\documentclass{optica-article}

\journal{opticajournal} 

\articletype{Research Article}
\usepackage{cite}
\usepackage[normalem]{ulem} 
\usepackage{amsmath,amsfonts}
\usepackage{graphicx}
\usepackage{textcomp}
\usepackage{xcolor}
\usepackage{physics}
\usepackage{fancyhdr}
\usepackage{braket}
\usepackage{comment}
\usepackage{subcaption}
\usepackage{lineno}
\newcommand{\AR}[1]{{\color{black}#1}}
\newcommand{\ARR}[1]{{\color{black}#1}}

\begin{document}

\renewcommand{\sout}[1]{}
\title{Analysis of polarization drift of optical signals over deployed aerial-inground fiber connections}

\author{Aneesh Ramaswamy\authormark{1,*}, Nageswara S. V. Rao\authormark{1},\\ Joseph C. Chapman\authormark{1}, Muneer Alshowkan\authormark{1}}

\address{\authormark{1}Computational Sciences and Engineering Division, Oak Ridge National Laboratory, Oak Ridge, TN, USA}
\email{\authormark{*}ramaswamya@ornl.gov} 


\begin{abstract*} 

Polarization \ARR{H-V} measurements of a classical 1550-nm signal are collected, and analyzed on 15-km hybrid aerial-inground fiber connections over 11 months. 
The spectral area and spectral moments of mHz-resolution Fast-Fourier-Transform (FFT) of these measurements are extracted, and related to temperature, humidity, wind speed, and time of day. Spectral area correlations show a strong diurnal structure: daytime maxima align with temperatures/wind speed peaks and humidity dips, with lower levels during the night. These diurnal patterns also show seasonality, with higher mean and variance in summer than winter. A random forest regressor is used to estimate FFT features from environmental measurements, informed by a theoretical model.

\end{abstract*}

\section{Introduction}
Polarization drift is a major challenge to assess and compensate for in current deployed fiber connections used for quantum communications. Variations in the fiber birefringence due to environmental factors, including temperature, humidity, and wind, are among the strongest sources of polarization drift. Such drifts can be highly stochastic and non-stationary~\cite{KarlssonPMD}, and consequently lead to difficulties in identifying their dependencies on environmental variables and in distinguishing noise. {Polarization drift characterization} \sout{The characterization of polarization drift would yield bounds on features such as the spectral intensity and center-of-mass frequency, and would}would \AR{enable} improved link-level state characterization, improved data transfer rates, and error mitigation. Several long-term studies \sout{measured} investigated polarization drift on various fiber infrastructure over long time periods, including over inground fiber~\cite{Woodwardfiber, Brodsky_2006_PMD}, aerial fiber~\cite{KhouloudNokiaANN, Bohata}, and hybrid aerial-inground fiber~\cite{Tremblay}. \sout{Research has also been conducted on the role of polarization drift in fiber sensing
}In addition, theoretical and simulation studies on state-of-polarization (SOP) fiber models~\cite{Poole:91, Wai:94bireffiber, czegledi2016polarization, bifrost} contribute to the development of \sout{practical models of polarization drift in fiber}\ARR{practical models of polarization drift, including digital twins, on fiber connections}. 

In classical multi-terabit communications employing polarization multiplexing, \sout{effects such as }polarization-mode dispersion (PMD) and SOP drifts are largely mitigated using polarization-maintaining fibers and digital signal processing (DSP)~\cite{app9061178, UlrichPM, Martinelli:06}.
These techniques are not as reliable for polarization-based quantum communication links on deployed fiber infrastructure\sout{, which is the infrastructure that quantum networks are generally expected to operate on}. Precise sub-second polarization stabilization is required in these networks for sustained high quantum entanglement throughput~\cite{rao_2025_ethcap} and fidelity~\cite{Craddock24Qu, Chapman:24}.
Characterization and estimation of signal parameters\sout{ (e.g., spectral moments)} based on environmental measurements \sout{of fiber infrastructure }can benefit link state monitoring and control over such quantum networks.

Polarization drift in both aerial and buried fibers is typically highly noisy and non-stationary, and environmental variables alone cannot reliably characterize drift without accounting for the fiber’s microscopic state. \ARR{Fiber optical properties vary through refractive-index modulation (e.g., thermo-optic) and photoelastic effects driven by mechanical perturbations such as acoustic vibrations, applied stress (twisting/bending) from wind and human activity, and strain from temperature and humidity gradients~\cite{Barcik:20}.
Altogether, these effects induce spatially uneven birefringence, resulting in differential group delay (DGD) and SOP rotations~\cite{wangphotonics10020103}.}
For the problem of highly noisy datasets with strong multivariate nonlinear dependencies and highly correlated features, machine learning (ML) methods\sout{ (such as random forests, support vector machines, and artificial neural networks}) can significantly improve estimation/prediction performance~\cite{Eastman:25, sena2025highfidelityquantumentanglementdistribution}. 
In particular, ensemble methods such as random forests \sout{perform well on problems with high amplitude multiplicative noise,}have performed effectively on datasets of polarization drift in aerial fibers~\cite{ProbstRFtuning, Biau2015ARF}. 
Previous ML-based approaches for SOP drift estimation in exclusively inground or aerial fiber have shown errors as low as $2.9\%$ (in mean average percentage error) using multiscale models trained on lagged data and weather~\cite{KhouloudNokiaANN}. However, such analyses have not been extended to connections composed of both fiber types over long periods spanning multiple seasons. \sout{Furthermore, past works focused primarily on forecasting rather than estimation using environmental measurements.}

\subsection*{Results overview}
In this work, we collected continuous 
\ARR{horizontal-vertical (H-V) basis} Stokes measurements of a 1550-nm continuous-wave signal
, and processed them using our multi-FFT signal characterization device
~\cite{Chapman:24}, for 11 months over hybrid aerial-inground fiber connections. 
Our analysis focuses on the FFTs spanning mHz to Hz frequencies. 
We extract global spectral features $\mathcal{Y}$, representing the spectral area, spectral centroid, spectral variance, spectral entropy, and the $\beta$-exponent of the $f^{-\beta}$ envelope. We use Spearman's correlation to characterize the relationship of $\mathcal{Y}$ with histories of environmental variables $\mathcal{X}$ representing the hour of day, temperature, relative humidity, wind speed and the changes of the latter three over 45-minute intervals. Also, we use random-forest regressors to estimate FFT features $\mathcal{Y}$ as functions of histories of the environmental variables $\mathcal{X}$. \sout{This estimation is informed by our theoretical model describing SOP drift over fiber.
We consider a detailed feature-importance analysis to be out of scope due to the high multi-colinearity of environmental variables over the day.}

We present four principal findings: (i) the spectral area generally obey diurnal cycles with larger averages and variances during daytime, and more regular cycles in spring, summer, and fall compared to winter; (ii) FFT characteristics of the signal imply strong multiplicative noise with resilient higher-order spectral moments and the signal is concentrated within 5.1\% of the 0.76 Hz FFT window; (iii) time of day and temperature show the strongest dependence on the spectral area, and wind speed shows the strongest dependence on the spectral centroid and $\beta$-exponent in winter; and (iv) ML estimators show distinct seasonal and daytime/nighttime patterns and estimation performance, and small overall gap (<2\% relative error) between test and train datasets \ARR{for the spectral area} under the root mean square logarithmic error (RMSLE).

\section{Methods}
\subsection{QNET testbed}
The QNET testbed provides twenty connections composed of aerial and inground fiber over network infrastructure at Oak Ridge National Laboratory~\cite{alshowkan21testbed}, see Fig. \ref{fig: Expsetup}. The aerial-inground fiber used for our experiments is an armored 48F, gel-free, single-tube, single-mode (OS2) ribbon cable with multiple water-swellable tapes.
A 1550 nm polarized signal (Pure Photonics PCL550) is transmitted through a Polatis all-optical switch into the fiber connections. \sout{Previous experiments have made use of hybridized connections consisting of subconnections of aerial-inground loops, and subconnections of fiber spools arranged in a telescoping design, provisioning entanglement distribution connections
.}

\subsection{Experimental design}
The setup for measurements of signal intensity of a polarized signal is shown in Fig. \ref{fig: Expsetup}, contained in 3U rack-mounted box. The experimental design in Appendix F of a prior work~\cite{Chapman:24} is replicated in our setup.
A continuous-wave 1550 nm signal is produced by a 10-kHz linewidth, polarized laser (Pure Photonics PCL550) and transmitted over aerial-inground fiber connection via an all-optical switch (Polatis). Power meters are situated before and after transmission through the fiber connection, to monitor power fluctuations in input laser. 
The polarization signal is measured (in $\mu$W) using a fiber polarization beam splitter (Thorlabs PBC1550SM-FC) that is configured to pass the horizontal (H) and vertical (V) polarized signals to two optical power meters (Thorlabs PM101A, S154C), resulting in two analog outputs $a$ and $b$. \ARR{By design, a partial representation of the polarization drift in the fiber (along the H-V axis) is captured, and similar drift dynamics are expected for other polarized intensity components.}
The two transmitted analog power outputs are sampled at 100 MHz by an analog-to-digital converter (ADC, Analog Devices AD9254), resulting in samples $A_{100M}$ and $B_{100M}$, and then 
transmitted to a field-programmable gate array with system on chip (FPGA-SoC, demonstration board Terasic ADCSoC with co-located Intel Cyclone V SoC hard processor). 

Further sample accumulation is carried out to obtain series ${A_j,B_j}$ for $j\in\{\text{mHz, Hz, kHZ, MHz}\}$, where samples at lower rates are obtained by averaging samples from higher rates. \ARR{The H-V intensities} are calculated over 11 months, from December 2024 to November 2025, by taking the ratios $A_j/(A_j+B_j)$ and normalizing. \sout{These signals are dimensionless quantities that correspond to an affine transformation of the Stokes vector, $(1+S_1/S_0)$, where $S_1$ is the $H-V$ axis-oriented Stokes component and $S_0$ is the non-projected optical power.} 
Three FFTs \sout{(magnitudes only)}\ARR{magnitudes} with 2048 bins are calculated: $Y_{\text{MHz}}$ with a bin width of 381.47 Hz; $Y_{\text{kHz}}$ with a bin width of 0.3725 Hz; and $Y_{\text{Hz}}$ with a bin width of 0.378 mHz.

Temperature ($T$), relative humidity ($H$), and wind-speed ($W$), are measured at 2 minute intervals at a 15 m altitude, from a local weather monitoring station. 
These weather samples are averaged and time-matched with FFTs. 
Hour timestamps (Hr) are also passed as inputs.

\begin{figure}[htbp]
    \centering
    \includegraphics[width=0.9\linewidth]{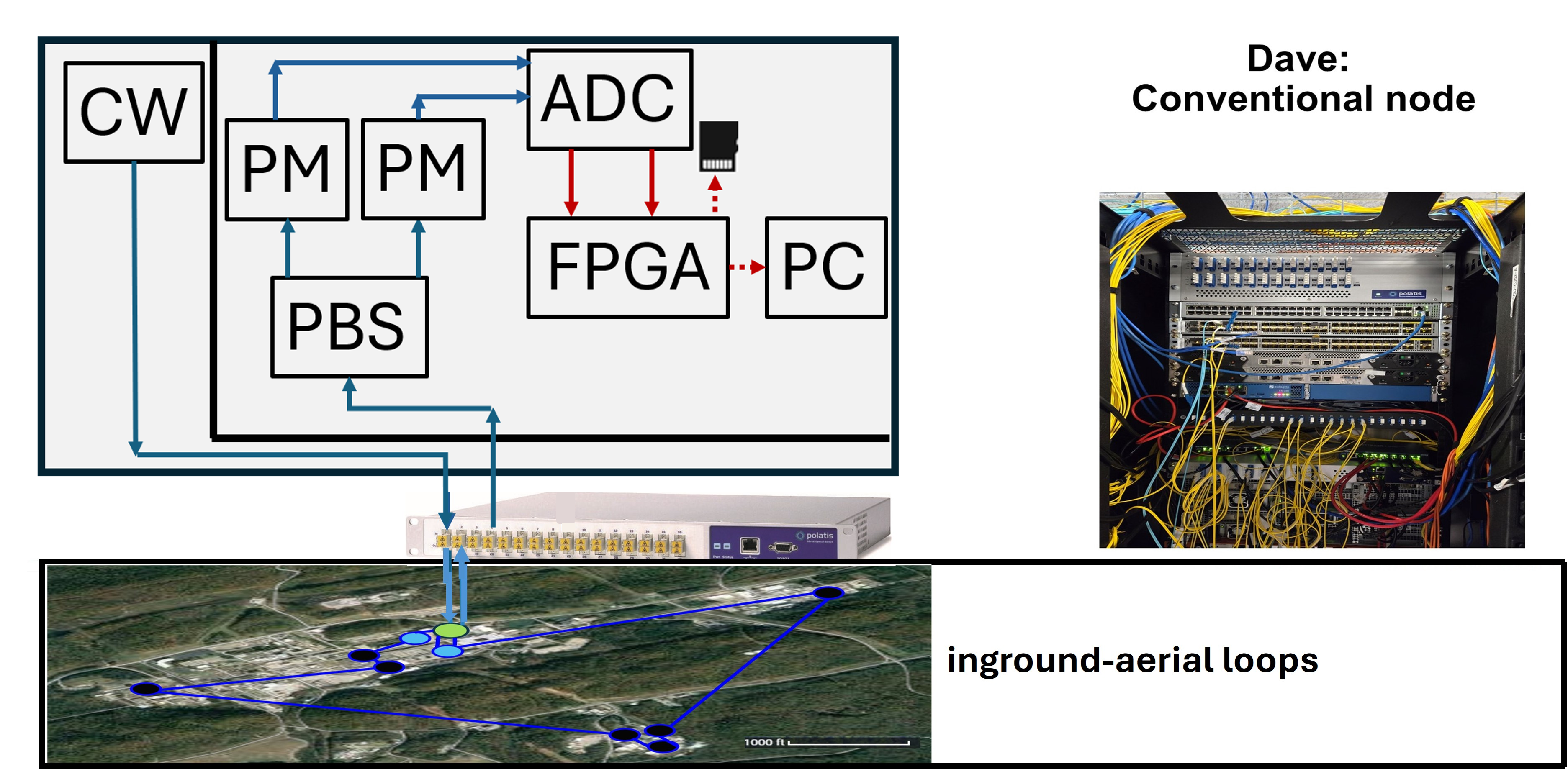}
    \caption{Polarized signal measurement and data collection setup (Top left inset). Map of the ORNL fiber infrastructure (bottom inset) that the laser traverses. The setup is housed in the server node Dave and the FPGA is connected to a Linux host. Blue lines: communication channels carrying the laser. Red lines: analog signals. Dotted lines: digital signals. CW: continuous-wave laser. PBS: Fiber polarization beam splitter. PM: optical power meter. ADC: analog-to-digital converter.}
    \label{fig: Expsetup}
\end{figure}

\subsection{Data processing}\label{sec: data_proc}
In this work, the focus is on the $Y_{{\text{Hz}}}$ dataset. MATLAB is used for importing the raw data, timestamp matching, and feature extraction (peak analysis). For the global features, the following are computed: the spectral area, $Y_{\text{Area}}[n]$ (the dimensionless sum of the first 40 FFT bins); the mean and uncertainty of the dimensionless $\beta$-parameter, $Y_{\beta}[n]$ and $Y_{\beta,\text{unc}}[n]$, respectively, that characterize the noise spectrum $f^{-\beta}$ (calculated from least-squares log-log fitting of FFT magnitudes over bins 5$\sim$100); the spectral centroid, $Y_{\text{sp,cen}}[n]$ (Hz); the spectral variance, $Y_{\text{sp,var}}[n]$ ($\text{Hz}^2$); and the spectral entropy, $Y_{\text{sp,ent}}[n]$ (dimensionless). The spectral area characterizes the signal power in a low frequency band (first 40 bins to exclude white noise). The spectral centroid and variance characterize the center of mass frequency and spread of the signal, respectively, while the spectral entropy characterizes the signal noisiness. These last three features are calculated over all 2048 FFT bins. The peak statistics in the FFT were also determined: the findpeaks function in MATLAB is used to find the peak heights of the first $10$ peaks in the FFT.

The remaining data analysis is done in Python. Correlation analysis using Pearson's correlation scores \cite{rodgers1988thirteen} is carried out to exclude redundant, strongly correlated features, and to obtain the final target feature set $\mathcal{Y}_{\text{target}}=\{Y_{\text{Area}},Y_{\text{sp,cen}},Y_{\text{sp,var}},Y_{\text{sp,ent}},Y_{\beta}\}$. \ARR{Plots of the dependencies between $Y$-features for summer and winter are in Supplemental 1.}  \AR{Principal component analysis (PCA) \cite{PCAreview} of the ML output variables is used to compute a reduced uncorrelated vector of PCA components (after feature normalization) that explains more than $95\%$ of target feature data variability, and the vector of PCA components is converted back to $\mathcal{Y}_{\text{target}}$.}

For the weather data, averages, $X_{k,\mu}$, and standard deviations, $X_{k,\sigma}$, are computed over 45-minute intervals for each input, $X\in\{T,H,W\}$. 
The change variable, $\Delta X_K$ (representing changes in the 45-minute interval values for $X\in\{T,H,W\}$) is also passed. 
The final input feature set passed is $\mathcal{X}=\{\text{Hr},T,H,W,\Delta T, \Delta H, \Delta W\}$. A Spearman's correlation analysis is used to quantify the monotonic relations between $\mathcal{X}$ and $\mathcal{Y}$ and distinguish between summer and winter trends. 
For quantifying the non-monotonic relation of $Y_k$ with Hr, the mutual information score (MI) is estimated using the K-nearest neighbor implementation in scikit-learn~\cite{scikit-learn}. We excluded the peak attributes (peak widths, frequencies, and quality factors) because Spearman's correlation scoring showed low dependencies (magnitudes $<0.1$) with all environmental inputs, as well as pronounced jitter, \ARR{see Supplemental 1}.

\subsection{\ARR{Physical model}}\label{sec: phys_model}

In this section, \sout{ we describe a general physical model of the polarization dynamics of an optical signal and establish connections to the environmental variables.}\ARR{we describe the physical dynamics for relating environmental measurements to polarization-drift spectral features, and for including measurement histories to develop estimators that account for delayed fiber responses to environmental measurements.}

Assuming time-independence, the polarization state of an optical signal, with constant intensity $S_0$, propagating along a fiber is described by the projected Stokes vector, $\vec{S}(z)=(S_1(z),S_2(z),S_3(z))^T$, where  \ARR{$S_j(z)$ represents the signal's relative intensity along a specific polarization axis on the Poincare sphere (horizontal/vertical, diagonal/anti-diagonal, and left/right -circular)}, at position $z$ on the fiber. The measured signal powers, $A$ and $B$, are related to the H-V polarized Stokes component, $(1+S_1/S_0)$, with the equation of motion given by~\cite{Poole:91},
\begin{equation}
    \partial_z\vec{S}=\zeta(z)\times \vec{S}(z)\label{eq: Stokes}
\end{equation}
where $\zeta(z)$ is the spatially local birefringence vector that acts as the generator of polarization drift. Integration of Eq. \eqref{eq: Stokes} yields a 3D rotation unitary matrix operating on the initial Stokes vector. \ARR{Its components represent a rotation about a specific axes on the Poincare sphere, driven by environmentally induced modulation of the medium's refractive indices.} These include non-mechanical effects (e.g., thermo-optic effects) and mechanical (localized twisting/bending of the fiber induced by thermal/hygroscopic expansion, and wind-driven motions). \ARR{In addition, human activities corresponding to heavy vehicular traffic and machinery can also induce stress-driven birefringence rotations. These activities generally induce abrupt spikes in the polarization drift on faster timescales than weather changes. Our testbed is situated in an area with little road activity and we do not monitor human activities. Hence, we exclude consideration of these activites in this work. For a more rigorous model of the polarization drift dynamics using a discretized fiber model, see Supplemental S1.}

\ARR{The fiber’s response to ambient conditions exhibits a time delay due to heat transfer mechanisms and mechanical relaxation. Developing accurate regression estimators thereby requires replacing the input environmental measurements with a recent history. Next, a brief description of the mathematical dependencies of the estimator on the inputs is given.}

\subsubsection*{Developing an estimator using environmental measurements as inputs}
A probabilistic map for the empirical time-domain signal, $\Xi:\{\tilde{X}_{k,f}\}\rightarrow (1+S_1/S_0)$, is constructed. \sout{between $\{\tilde{X}_{k,f}\}$ and the time-discretized relative intensity $(1+S_1/S_0)$.} 
An FFT is applied to the outputs of $\Xi$, yielding a vector output $\tilde{\Xi}$ that describes the probability distribution of FFT bins $Y_{\text{Hz}}$, given input $\{\tilde{X}_{k,f}\}$. 
Thus, the probabilistic map for the empirical data, $\Xi^{\prime}: \{\tilde{X}_{k,f}\}\rightarrow\mathcal{Y}$, is obtained. 
The ML estimator corresponds to the conditional expectation of this map obtained using measurements of environmental variables $\mathcal{X}$. 
A linear functional for the fiber variables, $\tilde{X}_{j,f}(t)=\int_0^t d\tau\text{ }\vec{G}_j(\tau)^T\vec{X}(t-\tau)$, can be found, with a response kernel, $\vec{G}_j(\tau)$, that is a property of the fiber, surrounding material, environment, and heat transfer methods.
Hence, multiple time-delayed measurements $X_j(t-\tau)$ are passed as inputs, as shown in the next section.

\subsection{Machine-learning methods}\label{sec: ML_methods}
With insights from theory, multi-output estimators are developed for all targets $\mathcal{Y}$ as a function of histories $\mathcal{G}_{n,\tau}[\mathcal{X}]$ of environmental parameters $\mathcal{X}$ over the past $n$ measurement separated by $\tau$ hours. 
\ARR{The objective is to construct an estimator} \sout{corresponds to}\ARR{that learns the regression function $E[Y_k|\mathcal{G}_{n,\tau}[\mathcal{X}]]$, $\forall Y_k\in\mathcal{Y}$, obtained based on measurements. A grid search was conducted on $(\tau,d\tau)$, where $\tau$ is the history length and $d\tau$ is the lag increment, to determine the history that minimized the training error.}
\sout{We exclude the time index $t$. 
This general description in terms of the history of inputs is to account for the equilibration time of the ambient fiber conditions detailed in the previous section.}
\AR{The target empirical output $Y_k\in\mathcal{Y}$ is considered to be a nonstationary signal with an unknown noise distribution. 
\sout{The goal is estimation of the deterministic component $E[Y_k|\mathcal{G}_{n,\tau}[\mathcal{X}]]= g_k(\mathcal{G}_{n,\tau}[\mathcal{X}])$ as a function of the input $\mathcal{G}_{n,\tau}[\mathcal{X}]$.}

A machine learning regression pipeline is \sout{constructed}{developed} for estimating the conditional mean for targets $\hat{Y}_k$. The initial data preprocessing is carried out as described in Section \ref{sec: data_proc}, and  random forest regressors are used for the ML models.
\sout{Random forests are chosen, owing to their reliable performance and ease of tuning for very noisy, small datasets, particularly with the use of bootstrapping
.}Random forests are chosen for their known robust performance with small, noisy datasets~\cite{ProbstRFtuning}.
The multi-output RandomForestRegressor function from the scikit-learn package is used ~\cite{scikit-learn}, where the PCA components (calculated from normalized $\mathcal{Y}$) are passed as targets, and $\mathcal{X}$ are passed as inputs. Lastly, the inverse PCA transform is applied to return the outputs for our machine learning pipeline, $\mathcal{Y}$.}

The training/validation dataset spans December 18th, 2024 to August 31st 2025 and the test dataset spans September 1st, 2025 to November 20th, 2025. 
For the inputs, two histories are considered:  (i) present environmental measurements $\mathcal{G}_{0,0}[\mathcal{X}]=\mathcal{X}$ and (ii) 4-hour spaced, 1 day measurements $\mathcal{F}_X=\mathcal{G}_{6,4}[\mathcal{X}]$. The second configuration accounts for the delayed response of fiber to ambient conditions. \sout{(such as temperature and relative humidity) due to the time required for dirt, sleet and other material on the fiber to equilibrate with the environment. Based on training RMSLE errors for different values of $(n,\tau)$, $\mathcal{F}_X=\mathcal{G}_{6,4}[\mathcal{X}]$ is chosen.}

For our random forest estimator trained on unlagged data ($\mathcal{X},\mathcal{Y}$) and lagged data ($\mathcal{F}_{\mathcal{X}},\mathcal{Y}$), the following parameters are used: a minimum leaf size of 10-15; a maximum tree depth of 10-15; a number of trees of 1000-2000; a maximum training sample size per tree of 0.15-0.25; and the root mean square error as the loss function~\cite{ProbstRFtuning}. For parameter fine-tuning, ten-fold shuffled cross-validation on the training/validation dataset is used. The choice of parameters and low complexity of the estimator was motivated by best practices in reducing estimator bias while preventing overfitting to noise, accounting for the high noise and structured variability in our dataset~\cite{LOPEZ2026344838, hastie09elements}.  
Final performance is reported based on RMSLE scores on the held-out test period, with the RMSLE error between the estimator $\hat{Y}$ and the empirical data $Y$ on the dataset $\mathcal{D}$ given by,
\begin{equation}
    \text{RMSLE}_{\mathcal{D}}(\hat{Y},Y)=\sqrt{|\mathcal{D}|^{-1}\sum_{(x,y)\in\mathcal{D}}\left(\log(1+\hat{Y}(x))-\log(1+y)\right)^2} .\label{eq: rmsle}
\end{equation}
RMSLE values were calculated using the five target features in $\mathcal{Y}$, which are all non-negative.%

Two baseline estimators are also generated to compare performance. A random forest estimator $\hat{Y}_{\text{base, Hr}}$ is trained using only the $X_{\text{Hr}}$ input feature, generalizing the naive mean baseline estimator. 
A second product estimator, $\hat{Y}_{\text{base, Hr+MA}}$, composed of two sub-estimators $\hat{Y}_{\text{base, Hr+MA},1}$ and $\hat{Y}_{\text{base, Hr+MA},2}$ is also developed to separately estimate the moving average and periodic component, respectively, of the spectral area and then combine the outputs. $\hat{Y}_{\text{base, Hr+MA},1}$ is a regularized linear regressor (using scikit-learn's Ridge Regressor) that was trained on the moving average (MA), $Y_{\text{Area,MA}}$. $\hat{Y}_{\text{base, Hr+MA},2}$ is a random forest regressor that was trained on the seasonal component. Both of these sub-estimators have input $X_{\text{Hr}}$. 
{Additionally, the cross-validation (CV) RMSE score is also computed to describe training performance. It is computed using Eq. \eqref{eq: rmsle} on each fold for the shuffled ten-fold split on the training dataset.} 

We quantify the success of the estimators based on: 
(i) comparison of the RMSLE errors with baseline models passed only time of day data; (ii) reductions in test and training RMSLE errors upon introducing new features\sout{ (e.g. lagged data)}; and (iii) differences in RMSLE errors between training and test datasets to evaluate generalization.

\section{Analyses Results: Correlation and ML Estimation}
Our FFT analysis revealed that the dominant drifts are of low frequency components from $3\times10^{-4}$ to $10^{-2}$ Hz. \sout{We excluded the peak widths, frequencies, and quality factors because Spearman's correlation scoring showed low dependencies (magnitudes $<0.1$) with all environmental inputs, over all seasons. There is also pronounced jitter in these peak features, possibly due to spectral
329 leakage from aperiodic, stochastic polarization drift bursts in the time signal over the FFT
330 intervals, making them unreliable for spectral characterization.} 
{To summarize the spectral features}, the approximate ranges are: 
spectral centroid is $(3.2\pm0.9)\text{ mHz}$; the spectral variance is $(1270\pm20)\text{ mHz}^2$ (the standard deviation/square root variance is $\sim10\times$ the spectral centroid); and the spectral entropy is $0.92\pm0.07$. The signal energy is highly localized: within one spectral standard deviation, the signal occupies $5.1\%$ of the $0.76\text{ Hz}$ FFT range.

\sout{This altogether reveals the FFT is highly noise-dominated, as expected from the multiple aperiodic bursts of polarization drifts, seen in the time-domain signal, due to factors such as wind gusts. 
Additional plots of the moving average, variance and raw data for these target features are in Appendix A. We next characterize the diurnal and seasonal trends of sub-Hertz components of $\mathcal{Y}$ with respect to inputs $\mathcal{X}$, as discussed in Section \ref{sec: data_proc}, using a correlation and ML estimator analysis.}
\ARR{In the next section, we describe the diurnal and seasonal trends of the sub-Hz polarization drift using a correlation analysis.}

\subsection{Correlation analysis}
\begin{figure}[htbp]
     \begin{subfigure}[b]{0.48\columnwidth}
         \centering
        \includegraphics[width=1.0\linewidth]{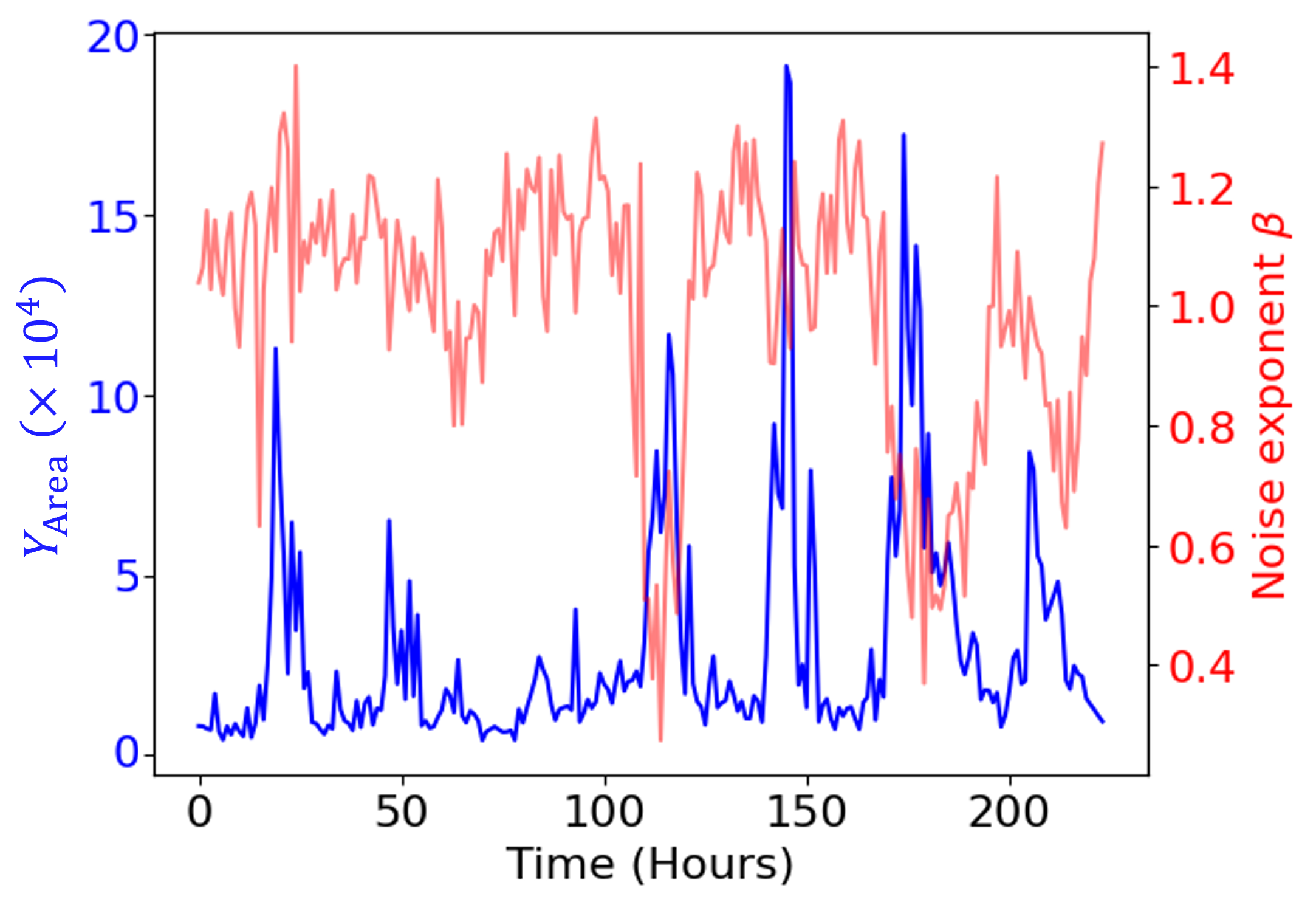}
        \caption{winter pol. drift}
     \end{subfigure}
     \hfill
     \begin{subfigure}[b]{0.48\columnwidth}
         \centering
        \includegraphics[width=1.0\linewidth]{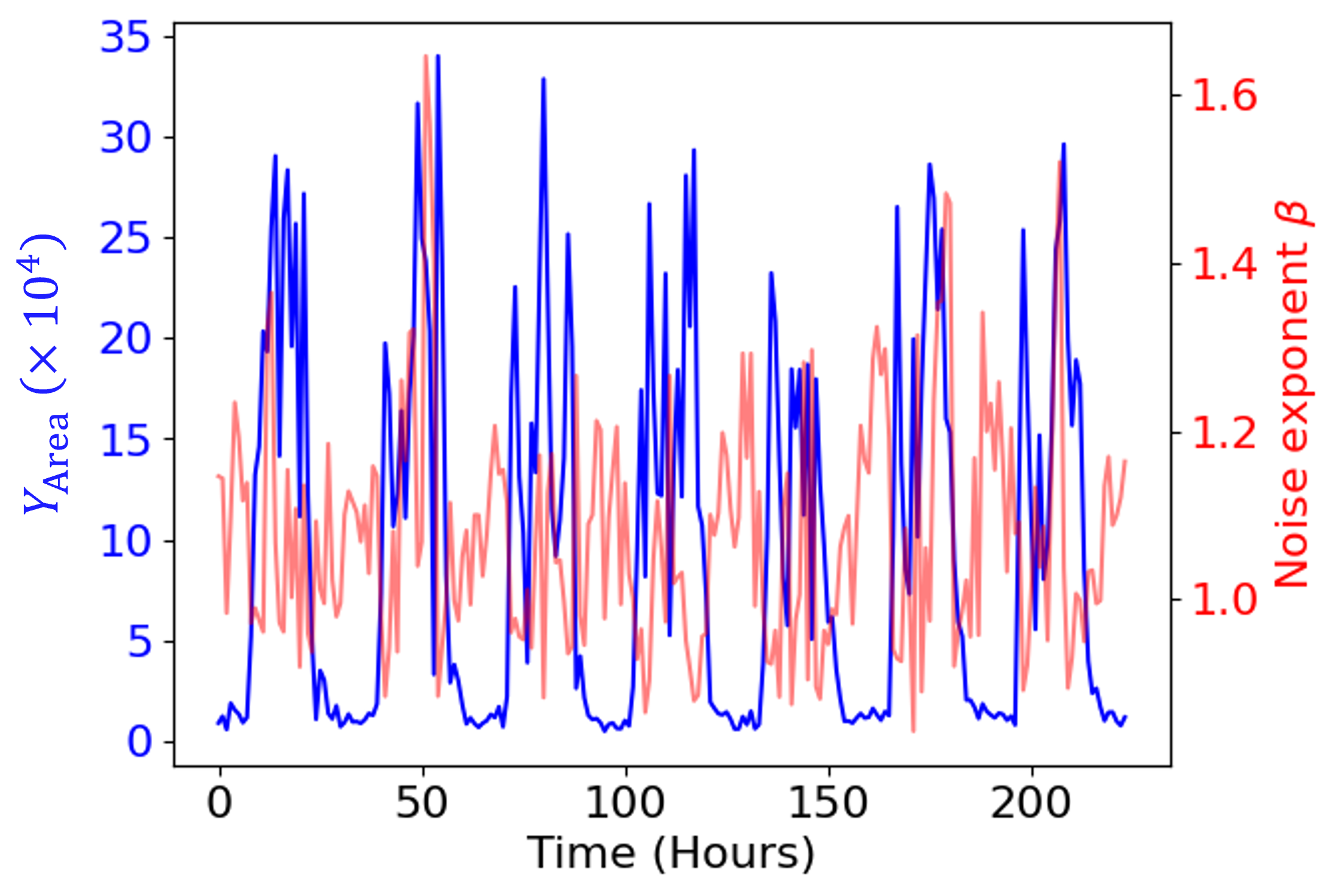}
        \caption{summer pol. drift}
     \end{subfigure}
     \hfill
     \begin{subfigure}[b]{0.48\columnwidth}
         \centering
        \includegraphics[width=0.93\linewidth]{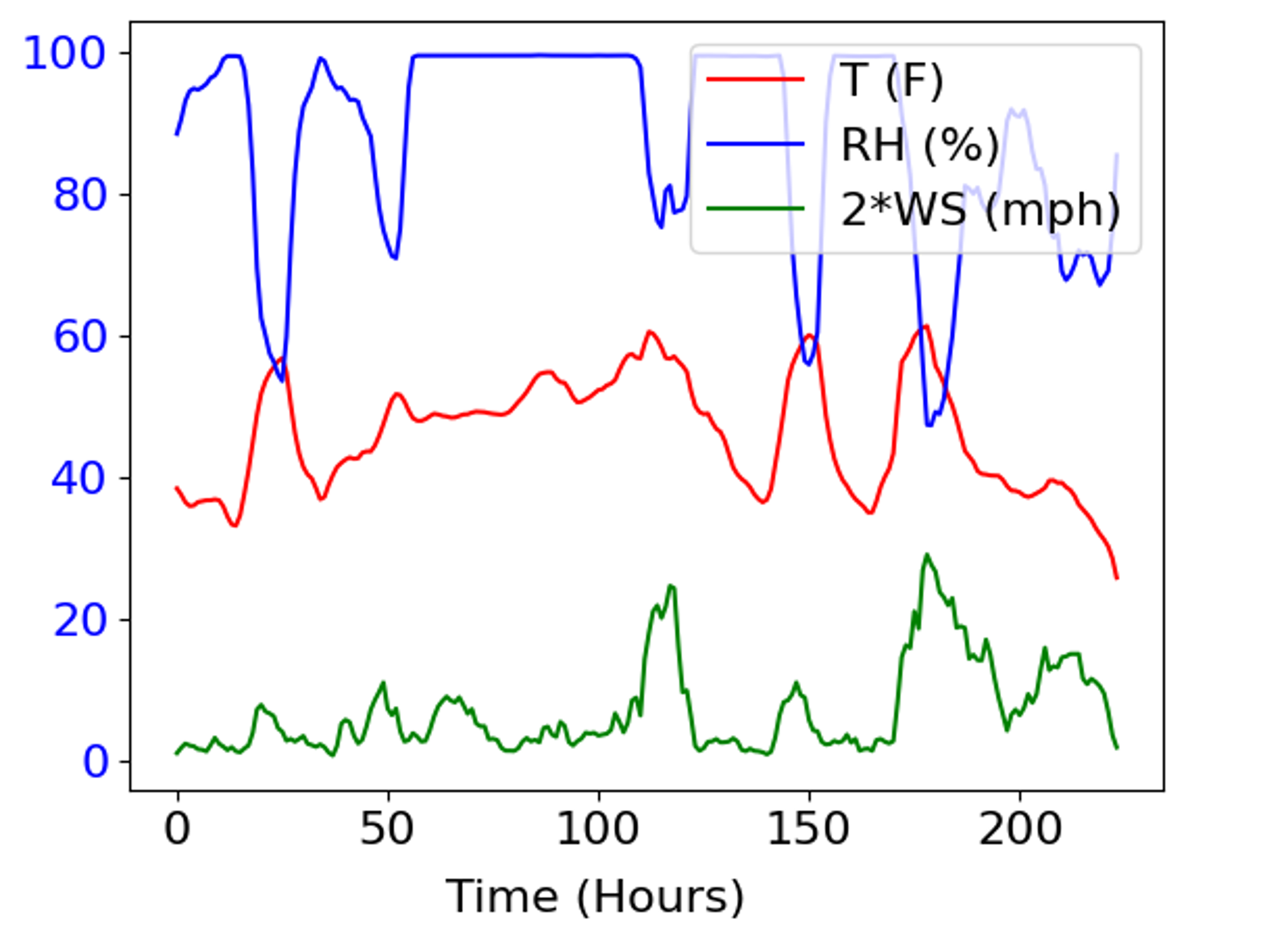}
        \caption{winter weather}
     \end{subfigure}
     \hfill
     \begin{subfigure}[b]{0.48\columnwidth}
         \centering
        \includegraphics[width=0.93\linewidth]{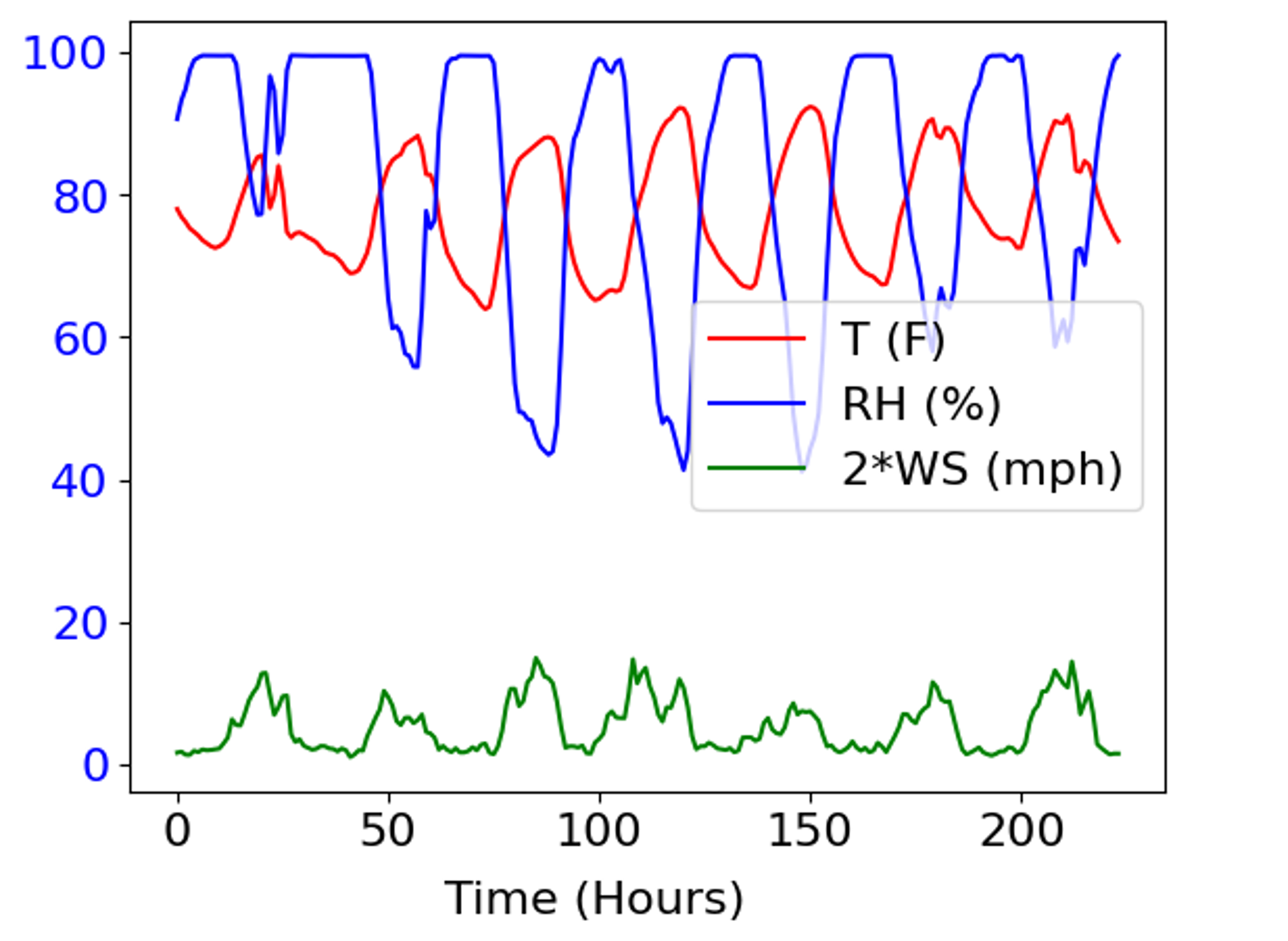}
        \caption{summer weather}
     \end{subfigure}
     \caption{A week of polarization drift in winter (a) and summer (b) showing the FFT spectral area $Y_{\text{Area}}$ and the beta-coefficient $\beta$. Corresponding weather measurements ($X=T,H,W$) are also shown for winter (c) and summer (d).}
     \label{fig: weekofdrift}
\end{figure}

We observed diurnal cycles that showed that the spectral area,  $Y_{\text{Area}}$, maximized in the afternoon during temperature peaks, wind speed peaks and humidity dips. Furthermore, one-week plots of $Y_{\text{Area}}$ showed flat troughs of minimal spectral area during nighttime and temperature lows (Fig. \ref{fig: weekofdrift}). 
\sout{Since hourly time has a non-monotonic and non-linear relation with the targets, Pearson's and Spearman's correlations are unreliable here to quantify the relation with time of day. We estimated the relation using the mutual information (MI) score, calculated using scikit-learn's K-nearest neighbor mutual regression score
.}
$X_{\text{Hr}}$ had an estimated MI of 0.57 in summer and 0.28 in winter, showing the higher regularity in summer. We caution that the estimated mutual information scores are less reliable indicators of relations than the Spearman's correlation and our use here is only to compare the contributions between winter and summer.

\begin{figure}[htbp]
     \begin{subfigure}[b]{0.47\columnwidth}
         \centering
        \includegraphics[width=1.0\linewidth]{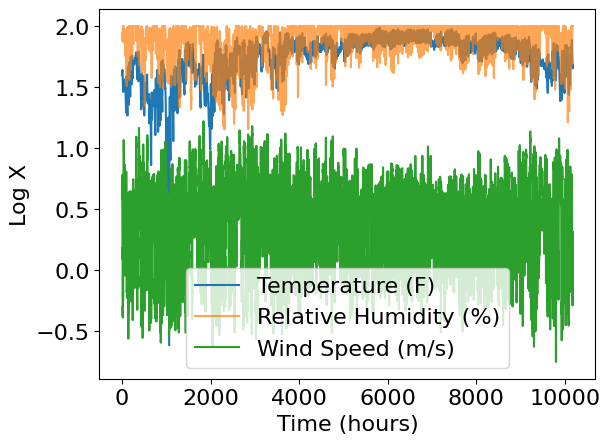}
        \caption{weather semilog plot}
     \end{subfigure}  
     \hfill
     \begin{subfigure}[b]{0.47\columnwidth}
         \centering
        \includegraphics[width=1.0\linewidth]{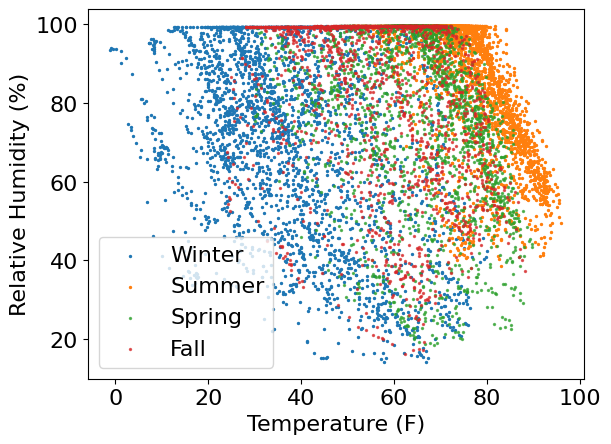}
        \caption{seasonal T-H plot}
     \end{subfigure}
     \hfill
     \begin{subfigure}[b]{0.48\columnwidth}
         \centering
         \includegraphics[width=1.0\linewidth]{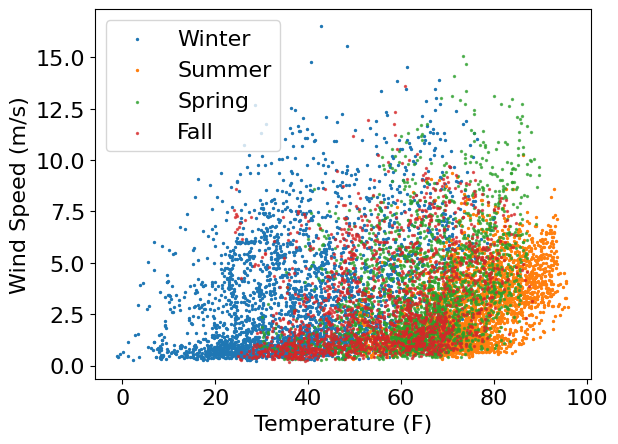}
         \caption{seasonal T-W plot}
     \end{subfigure}
     \hfill
     \begin{subfigure}[b]{0.47\columnwidth}
         \centering
         \includegraphics[width=1.0\linewidth]{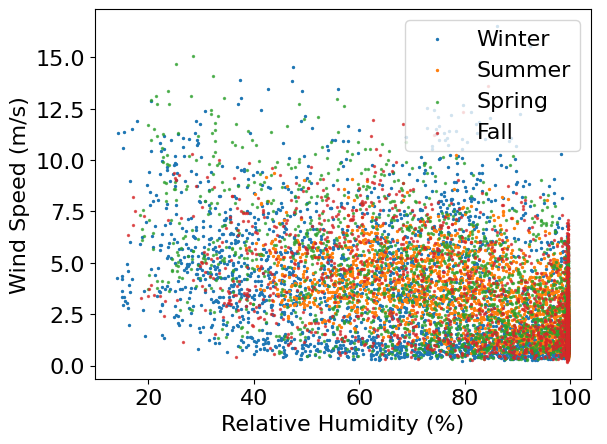}\caption{seasonal H-W plot}
     \end{subfigure}
     \caption{(a) Log plot of the exponentially weighted means of temperature, relative humidity, and wind speed measurements from 12/18/24 to 11/20/2025. (b-d) Weather feature space plots for different seasons.}
     \label{fig: weatherA}
\end{figure}

As for the longer term trends ($>1\text { day}$), we first look at the weather plots for the semi-log plots of the temperature, relative humidity and wind speed versus time and the two-feature plots (Fig.~\ref{fig: weatherA}). There was high seasonal variation in the range of weather conditions (b-d) with summer showing the least variation and winter the largest. The seasonal differences in weather variability was also reflected in the Pearson's correlation heatmaps of the environmental variables for summer and winter \sout{(Fig. 5)}, \ARR{see Supplemental 1}. 

\begin{figure}[htbp]
    \centering
    \begin{subfigure}[b]{0.46\columnwidth}
        \centering
        \includegraphics[width=\linewidth]{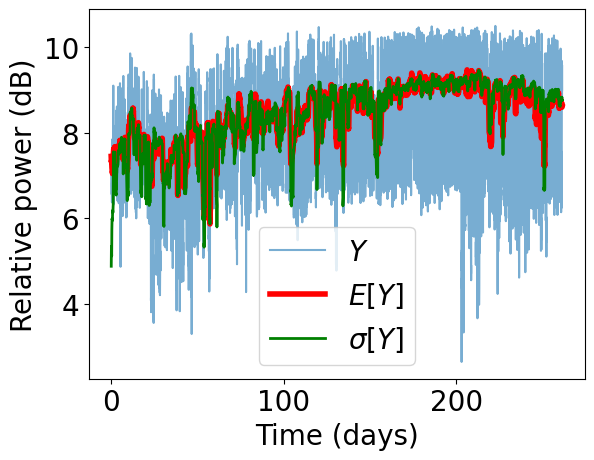} 
        \caption{versus time}
        \label{fig:411a}
    \end{subfigure}
    \hfill
    \begin{subfigure}[b]{0.48\columnwidth}
        \centering
        \includegraphics[width=\linewidth]{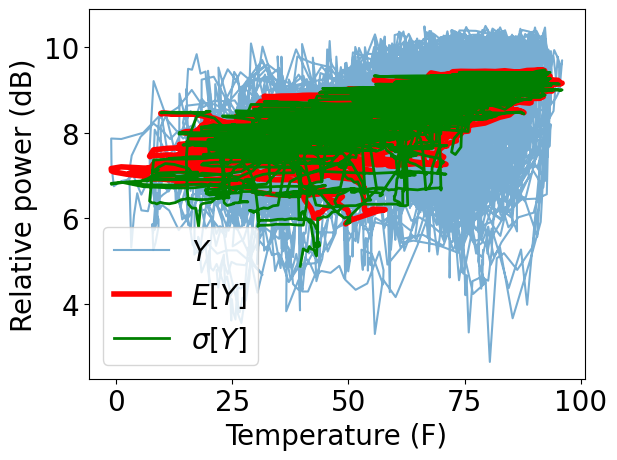} 
        \caption{versus temperature}
        \label{fig:411b}
    \end{subfigure}    
     \caption{(a) The spectral area, $Y=Y_{\text{Area}}$, shows clear signs of heteroskedasticity in the semilog plots versus time. (b) The positive temperature dependence is also made clear by the positive linear trend in semilog space. The moving average $E[Y]$ and rolling standard deviation $\sigma[Y]$ are plotted.}
     \label{fig: heteros}
\end{figure}

\sout{As a consequence of the high weather correlations in summer, $Y_{\text{Area}}$ has more regular cycles in summer than in winter, with close alignments of local extrema of $Y_{\text{Area}}$ with wind speed, temperature, and relative humidity extrema. Also, $Y_{\text{Area}}$ is maximized approximately at the point of peak average temperature, coinciding with mid-July (Fig. 6). It is readily apparent that both the average and variance of $Y_{\text{Area}}$ are related with increases in seasonal temperature, implying non-stationary variance (heteroskedasticity) of $Y_{\text{Area}}$ with respect to seasonal temperature, supporting the physical model detailed in Section 2.4. To analyze the long-term temperature-driven trends, we computed the moving average and variance of the spectral area over 1-day windows (Fig. 6). The physical model indicates that temperature directly changes the birefringence in the fiber, which along with our observations of the strong correlation with the temperature change, likely suggests that higher temperatures were making the fiber more vulnerable to fiber stresses that generate birefringence changes. We qualitatively observed a monotonic relation between temperature and the moving average of the spectral area (Fig. 6 b).}

\sout{To quantify the monotonic relations, we also computed the Spearman's correlation score between $X_k$ and $Y_k$, in summer and winter, to estimate the degree of monotonicity between inputs and targets (Fig. 7). The changes $\Delta T$, $\Delta H$, were the most correlated, and wind speed, $W$, had higher correlations for $Y_{\text{Area}}$, $Y_{\beta}$, and $Y_{\text{sp,cen}}$ in winter than in summer. In addition to the greater appearance of higher wind speeds ($>8\text{ mph}$) seen in winter as a cause, we speculate the presence of additional wind-induced fiber motion phenomena specific to winter: for example, close to zero and sub-zero temperatures are associated with galloping motions in ice-coated fibers that can result in long-period ($<1$ Hz) drift oscillations [29]. This is in opposition to summer wind-related fiber motions which are generally much higher frequency (e.g aeolian vibrations) [36].} 

\sout{Next, we consider the exponent $Y_{\beta}$ that characterizes the $f^{-\beta}$ profile of the FFT. On average, $Y_{\beta}$ hovered about $1.0\pm0.2$, implying presence of flicker noise, and showed more frequent dips in winter than summer. This could be explained by the decreases in $\beta$ with respect to large wind speeds ($>8\text{ mph}$), which are more common in winter. Additionally, the Spearman's correlation scores of $Y_{\beta}$ showed significantly more negative scores with wind speed/wind speed standard deviation in winter (-0.47,0.44) versus summer (-0.12,-0.09) suggesting there are winter-specific trends that cause the sub-Hz FFTs to flatten (become more like white noise). This was also accompanied by a higher Spearman's correlation score of $Y_{\text{sp,cen}}$ with wind speed/wind speed standard deviation in winter (0.64,0.60) versus summer (0.16,0.15), as flattening the $f^{-\beta}$ profile increases the center of mass frequency.}

\begin{figure}[htbp]
     \begin{subfigure}[b]{0.49\columnwidth}
         \centering
        \includegraphics[width=1.0\linewidth]{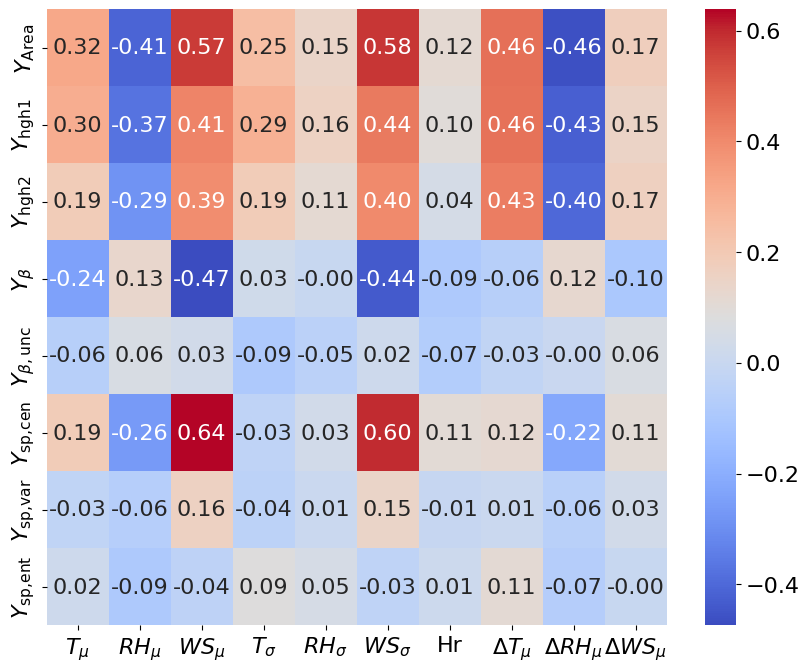}
        \caption{winter $XY$ correlations}
     \end{subfigure}
     \hfill
    \begin{subfigure}[b]{0.49\columnwidth}
         \centering
        \includegraphics[width=1.0\linewidth]{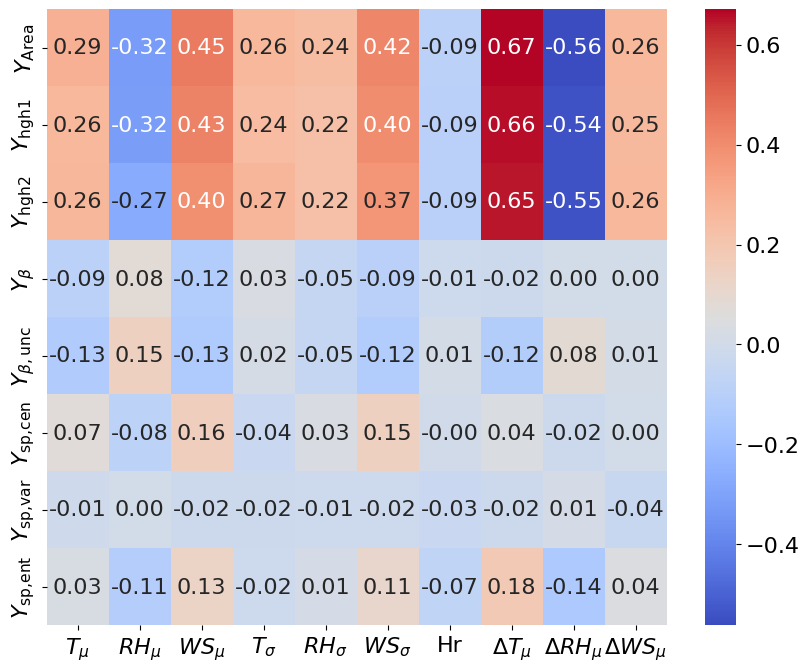}
        \caption{summer $XY$ correlations}
     \end{subfigure}
     \caption{Seasonal correlations of drift variables $Y_j$ versus weather variables $X\in\{T_r,H_r,W_r,\text{Hr}\}$ shown using calculations of Spearman's correlation.}
     \label{fig: weatherXYcorrelates}
\end{figure}

\ARR{Moving to the spectral feature analysis, the spectral area showed more regular cycles in summer than in winter. The computed 1-day moving average increased monotonically with temperature and peaked around mid-July, while the standard deviation exhibited similar relations with temperature, showing heteroskedasticity (Fig.~\ref{fig: heteros}).
Spearman correlations scores revealed strong associations between spectral area and  $\Delta T$ and $\Delta H$, with seasonal variation (Fig. \ref{fig: weatherXYcorrelates}). Wind speed exhibited stronger correlations with the spectral area, $\beta$, and the spectral centroid in winter than in summer. In particular, wind speed correlations with $\beta$ were more negative in winter than summer (-0.47 versus -0.12), while its correlation with the spectral centroid was more positive (0.64 versus 0.16). Wind-speed variance showed similar seasonal differences. Altogether, these correlations indicate a flatter spectrum under windier winter conditions.}

\sout{Thus observations so far are consistent with our physical model for the empirical data in Section \ref{sec: phys_model}. Spearman correlation analysis reveal monotonic dependencies on the environmental variables, $X$, which physically correspond to the environmental modulation of the slow-axis birefringence in the fiber and SOP drifts driven by uneven fiber expansion. Finally, there is an increasing trend in the average magnitude of daytime polarization drifts (spectral area) with higher weekly temperatures. This behavior is theoretically justified by the reduction in slow-axis birefringence, $\gamma_T(T-T_0)$, with elevated temperatures, resulting in an increased effect of the stochastic drifts, $\delta\zeta_{t}$.} 
The diurnal variation in the spectral area suggests that the current \ARR{air} temperature and relative humidity \sout{were not necessarily the primary factors for the polarization rotation}\ARR{do not fully characterize polarization drift}, as similar temperatures and relative humidity during daytime and nighttime do not correspond to the same spectral areas\sout{values}. This suggests \sout{that the polarization drift has a more complex dependency}a dependence on $\Delta T$ and $\Delta H$. \sout{with heating/humidity drops (daytime)) indicating an agitation (increased uneven stresses) and cooling/humidity peaks (nighttime) indicating a relaxation. As heating is driven by both solar radiation and convection with the surrounding air (nighttime cooling is driven by convection only), the spectral area during daytime is expected to be be significantly higher, due to the higher induced $\partial_tT_f$ in the fiber, than the spectral area during nighttime. This was shown in Eq. (4) and is supported by our results.} \ARR{It's likely that daytime heating and humidity decreases induce uneven fiber stresses, while nighttime cooling and humidity increases generate stress relaxation. Solar heating can also heat the fiber temperature above the ambient air temperatures, potentially leading to the prominent daytime spectral area peaks. These observations are consistent with polarization drift generated by large stresses induced by fast thermal expansion through the stress-optical effect (see Supplemental 1 for a brief discussion).}

The \sout{other}remaining target features, $Y_{\text{sp,var}}$ and $Y_{\text{sp,ent}}$, showed small difference in the Spearman's correlation scores over winter and summer, reflecting the lack of seasonal variation of these signal characteristics. The high spectral entropy, $Y_{\text{sp,ent}}\geq0.9$, and the temperature-dependent standard deviation in $Y_{\text{Area}}$ indicate a multiplicative noise component consistent with \sout{our theoretical framework}\ARR{the dynamical evolution of polarization drift (Eq. \eqref{eq: Stokes}), specifically the random polarization drifts caused by bends/twists and uneven expansion in realistic fiber}. 
\ARR{In the next section, we describe the results of our machine learning estimation analysis.}

\subsection{ML Estimator results}
We train random forests to estimate $\hat{Y}$ of target spectral features, 
$\mathcal{Y}$, as a function of different histories of the environmental inputs $\mathcal{G}_{n,\tau}[\mathcal{X}]$. The objective here is to show how ML approaches can be used to estimate the deterministic polarization drift behavior throughout the whole fiber, \ARR{when we do not have access to local measurements along fiber segments that would enable us to use} \sout{the theoretical fiber}\ARR{deterministic multi-segment models} \sout{detailed in the}\ARR{(see Supplemental 1)}. Calculations of the RMSLE revealed large proportional training errors for $\hat{Y}_{\text{Area}}$ that we determined to be not reducible with the use of more complex estimators without overfitting to noise.
\begin{figure}[htbp]
    \centering
    \begin{subfigure}[b]{0.48\columnwidth}
        \centering
        \includegraphics[width=\linewidth]{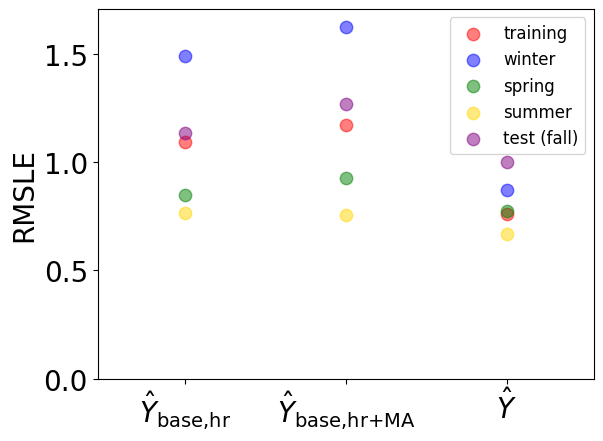} 
        \caption{daytime}
        \label{fig: 721a}
    \end{subfigure}
    \hfill
    \begin{subfigure}[b]{0.48\columnwidth}
        \centering
        \includegraphics[width=\linewidth]{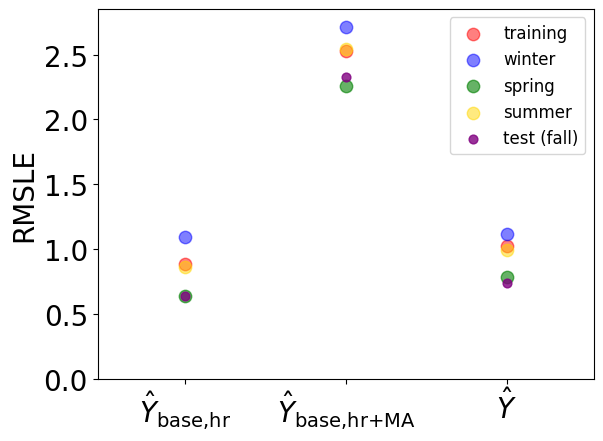} 
        \caption{nighttime}
        \label{fig:721b}
    \end{subfigure}    
     \caption{We compare the RMSLE scores for the different estimators for the spectral area $Y_{\text{Area}}$ trained on the unlagged dataset $(X,Y)$, with scores evaluated during daytime (8 AM to 4 PM), and nighttime (7 PM to 6 AM). The RMSLE training and the test (fall) scores are given, as well as the scores on each season.}
     \label{fig: baseline}
\end{figure}
\sout{To justify that the random forest estimator, $\hat{Y}=\hat{Y}_{\text{Area}}$, trained on unlagged data, $(X,Y)$, has learned dependencies with weather, we compared the errors for the spectral area target with the errors for the two baseline estimators, $\hat{Y}_{\text{base, hr}}$ and $\hat{Y}_{\text{base, hr+MA}}$. Fig. \ref{fig: baseline} shows a comparison of the training and seasonal folds (additionally split over daytime and nighttime hours) RMSLE values for the random forest and baseline estimators, detailed in Section \ref{sec: ML_methods}. During daytime, $\hat{Y}_{\text{base, hr}}$ failed to be as accurate as $\hat{Y}$, especially for winter and fall. For nighttime, however, $\hat{Y}_{\text{base, Hr}}$ performed surprisingly well in spring, summer and fall (test), showing generally better training and seasonal fold scores than $\hat{Y}$.}

\sout{At the same time, we clearly see that the inclusion of the additional environmental feature inputs corresponds to improved daytime estimation of the spectral area. Additional data coverage can resolve the nighttime underestimation. We see that $\hat{Y}_{\text{base, hr}}$ has not learned the seasonal trends well as it performs much worse in winter, owing to hourly time not being a good indicator of the winter weather patterns as compared to other seasons (as the weather correlation plots in Fig. \ref{fig: weathercorrelates} show). The combination baseline estimator $\hat{Y}_{\text{base, hr+MA}}$ performed the worst out of the three (especially at nighttime) and it is unclear whether this is an issue with the use of linear regressors to estimate the moving average, or an issue of training on a ratio dataset, $Y_{\text{Area}}/Y_{\text{Area,MA}}$. It is clear that our estimator $\hat{Y}$ delivered the most consistent performance on all folds, significantly reducing the error in winter. We also saw that $\hat{Y}$ during the summer and spring had the smallest RMSLEs for all estimators during daytime; and spring and fall had the smallest RMSLEs during nighttime. We note that the daytime trends of the fall test dataset were harder to estimate (the fall RMSLE score is higher than all other seasons even when included in training). This could be a result of unlearned weather patterns specific to fall that cannot be generalized from other seasons or unlearned trends due to underfitting, or fundamental differences in seasonal noise distributions. }

\ARR{Figure \ref{fig: baseline} compares the spectral-area estimators from the ML model with the two baseline estimators using the RMSLE metric, as detailed in Section \ref{sec: ML_methods}. Weather inputs reduced daytime errors, particularly in winter, with daytime errors largest in fall. However, the hour-only random forest baseline performed better at night for spring, summer, and fall, indicating that additional weather inputs did not consistently improve estimation performance.} 
This suggests that the presence of additional inputs is worsening the nighttime RMSLE error for $\hat{Y}_{\text{Area}}$ for spring, summer, and fall. \ARR{The combined moving-average and periodic baseline showed the worst performance with the highest RMSLE, suggesting other approaches may be needed for seasonal-trend decompositions.}

\sout{To account for the delayed response for the fiber to respond to the measured environment, we conducted a grid search on $(\tau,d\tau)$, where $\tau$ is the history length and $d\tau$ is the lag increment, to determine the history that minimized the RMSLE. Using the spectral area, $Y_{\text{Area}}$, as the output with RF estimators, we determined that $({\tau},d\tau)=(24,4)$ hours yielded a $5.4\%$ relative reduction in the test RMSLE, at 0.888, and a training CV RMSLE of 0.887. Therefore, we trained an additional estimator on a one day, 4-hour spaced, history $\mathcal{F}_X$.
Table \ref{table2} shows the CV and test RMSLE scores of the target outputs for the unlagged $(X,Y)$ and lagged $(\mathcal{F}_X,Y)$. Estimated target features apart from $\hat{Y}_{\text{Area}}$ and $\hat{Y}_{\beta}$, showed negligible change in the RMSLE scores with historical inputs.
The spectral centroid, $Y_{\text{sp,cen}}$, showed a CV and test RMSLE error (approximates to RMSE for small magnitudes) that was approximately twice the FFT bin spacing ($0.37\text{  mHz}$) suggesting that the time-frequency resolution was not the fundamental bottleneck.}

\begin{figure}[htbp]
    \centering
    \begin{subfigure}[b]{0.48\columnwidth}
        \centering
        \includegraphics[width=\linewidth]{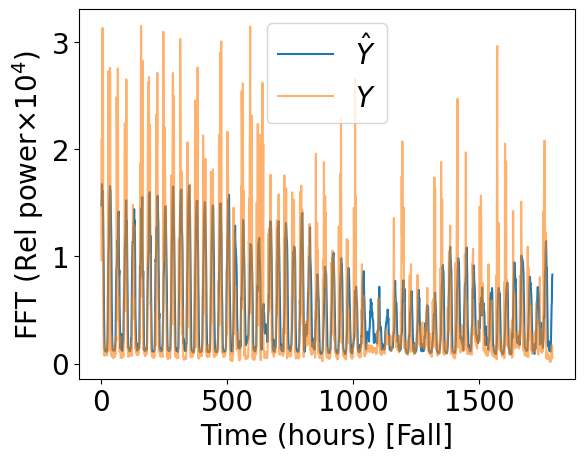} 
        \caption{Sep 1st to Nov 20th}
        \label{fig:521a}
    \end{subfigure}
    \hfill
    \begin{subfigure}[b]{0.48\columnwidth}
        \centering
        \includegraphics[width=\linewidth]{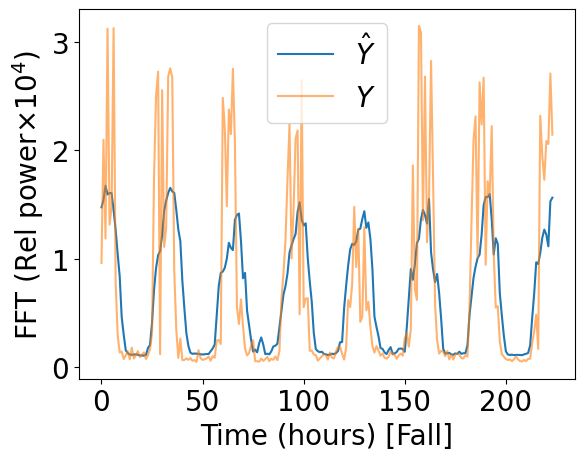} 
        \caption{week of Sep 1st-7th}
        \label{fig:521b}
    \end{subfigure}    
     \caption{Plots of the spectral area showing the empirical data, $Y$, and results from the random forest estimator trained on $\mathcal{F}_X$, $\hat{Y}_{\text{Area}}$ for the test dataset (fall). Plots over the full period (a) and a weekly period (b) reveal the random forest can effectively capture the diurnal patterns and overall trend but not the exact amplitudes.}
     \label{fig: RF_test}
\end{figure}

\ARR{Next, we considered the estimation performance with historical environmental measurements.} Including a 24-hour history of environmental measurements sampled at 4-hour intervals, $(\mathcal{F}_X,Y)$, reduced the spectral-area test RMSLE from 0.939 to 0.888 with a cross-validation RMSLE of 0.887 (Table \ref{table2}). Changes in test RMSLE for the other spectral features were negligible. The spectral centroid, $Y_{\text{sp,cen}}$, yielded a RMSLE that was approximately twice the FFT bin spacing ($0.37\text{  mHz}$) suggesting that the time-frequency resolution was not the fundamental bottleneck.
\ARR{The estimators reproduced the diurnal cycles and broader seasonal trend but did not fully capture individual peak and trough amplitudes or faster fluctuations (Fig. \ref{fig: RF_test}).}

\ARR{In the next section, we discuss the findings of our analyses in the context of the physical mechanisms for polarization drift in fibers, and address limitations in the methodology.}

\begin{table}[h!]
    \centering
    \small
    
    \begin{tabular}{|c|c|c|c|c|}
        \hline
        \textbf{Estimator} & \textbf{Dataset} & \textbf{CV RMSLE} & \textbf{CV} $\sqrt{\sigma_{MSLE}}$ & \textbf{test \textbf{RMSLE}}\\
        \hline
        $\hat{Y}_{\text{Area}}$ & $X$ & 0.959 & 0.022 & 0.939\\
        $\hat{Y}_{\text{Area}}$ & $\mathcal{F}_X$ & 0.887 & 0.036 & 0.888\\
        \hline
        $\hat{Y}_{\beta}$ & $X$ & 0.0741 & 0.0028 & 0.067\\
        $\hat{Y}_{\beta}$ & $\mathcal{F}_X$ & 0.0758 & 0.0018 & 0.066\\
        \hline
        $\hat{Y}_{\text{sp,cen}}$ & $X$ & 7.4E-4 & 2E-5 & 7.2E-4\\
        $\hat{Y}_{\text{sp,cen}}$ & $\mathcal{F}_X$ & 7.5E-4 & 2E-5 & 7.0E-4\\
        \hline
        $\hat{Y}_{\text{sp,var}}$ & $X$ & 1.1E-4 & 4E-6 & 1.1E-4\\
        $\hat{Y}_{\text{sp,var}}$ & $\mathcal{F}_X$ & 1.1E-4 & 3E-6 & 1.1E-4\\
        \hline
        $\hat{Y}_{\text{sp,ent}}$ & $X$ & 3.7E-2 & 7E-4 & 3.5E-2\\
        $\hat{Y}_{\text{sp,ent}}$ & $\mathcal{F}_X$ & 3.7E-2 & 1E-3 & 3.5E-2\\
        \hline
    \end{tabular}
    \caption{Cross-validation (CV) RMSLE scores and test RMSLE scores using the random forest estimator $\hat{Y}$ for different quantities. $\sigma_{MSLE}$: sd. of Mean square logarithmic error. $X$: Weather and hour variables, $\mathcal{F}_X$: History of $X$.}\label{table2}
\end{table}

\section{Discussion}
\sout{The results show a dominant diurnal trend in the spectral area, $Y_{\text{Area}}$. The presence of a nightimes drop throughout the year, regardless of temperature, humidity and wind speed, is consistent with previous works [18], and is indicative of the role of positive temperature gradients, negative humidity gradients, and wind speed peaks in determining polarization drift peaks, rather than average daily/nighttime readings. We observed that similar temperature readings during daytime and nighttime did not correspond to the same spectral area behavior (peak versus trough). We explained this as a result of the different heating mechanisms present during daytime (solar-radiative-heating and convention) and nighttime (convection). The spectral area of the drift peaked when $T$ reached a maximum, and $\Delta T$ became zero, which coincides with the peak daily solar intensity on non-cloudy days. There was a strong seasonal trend in the daytime maxima of the spectral area with larger values and more regular day-night cycles observed in summer than in winter. Furthermore, the moving average and variance of $Y_{\text{Area}}$ also revealed strong
heteroskedasticity in $Y_{\text{Area}}$ which coincided with the temperature moving average, supporting the role of temperature as a long-term indicator of drift amplitude and variance. Correlation analysis also revealed that there was a strong decreasing monotonic relation between wind speeds and the
$\beta$-exponent of the FFT profile ($f^{-\beta}$) in winter. While we expected wind speed to have a smaller effect for the sub-Hertz polarization drifts, winter conditions required more nuance. This is likely owing to the presence of galloping and other low frequency wind-induced fiber strain unique to winter [36].}

\ARR{The observed diurnal patterns in the spectral area, $Y_{\text{Area}}$ indicate that the measured air temperature is not a complete indicator of temperature-induced-polarization drift.} Similar temperatures during daytime and nighttime did not exhibit similar trends. Low drift activity during nighttime is consistent with previous works~\cite{Barcik:20}, and may reflect differences in heat transfer conditions, including solar heating and convention.
Increased heating and decreased humidity during the day could produce different fiber stresses from cooling and increased humidity at night.

Additionally, we observed a strong seasonal trend in the daytime maxima of the spectral area with larger values and more regular day-night cycles in summer than in winter. Correlation analysis also revealed that there was a strong decreasing monotonic relation between wind speeds and the $\beta$-exponent of the FFT profile ($f^{-\beta}$) in winter. This is likely owing to the presence of galloping and other low frequency wind-induced fiber strain unique to winter \cite{EXFO}. \ARR{These phenomena were not monitored in this work and can be a prospect for future study.} 

\sout{Our ML analysis showed that random forest estimators captured the weather dependencies with a <2\% relative error between test and training RMSLE scores (including both nighttime and daytime hours). While the absolute errors for the estimators, $\{Y_{k}\}$, were significant ($>100\%$ RMSE); we emphasize that the polarization drift contains high noise integrated over the 45-minute intervals, and our goal is not to predict the entire polarization signal but estimate the expectation of the weather dependent part of the target features, $E[Y_k|\mathcal{G}_{n,\tau}[\mathcal{X}]]$, for $Y_k\in\mathcal{Y}$.  We observed that the addition of lagged data, $\mathcal{F}_\mathcal{X}$, led to better estimation of the troughs at night in particular, and reduced sensitivity to the sub-peak features. This reduction in error is likely explained by the inclusion of lagged measurements that account for the time required for the fiber's ambient conditions to respond to the environmental changes, considering additional material that may be on the fiber that needs to first equilibrate with the environmental conditions (e.g dirt, sheath, fiber huts, sleet).}

\ARR{The random forest estimators captured broad weather-correlated patterns in the target variables}, showing a <2\% relative error between test and training RMSLE scores (including nighttime and daytime). However, substantial errors ($>100\%$ root-mean-squared error) were observed, \ARR{particularly with sub-peak features with durations shorter than the timescales of weather variations.} 
The limited tree complexity, see Sec. \ref{sec: ML_methods}, may have smoothed both noise and weather-dependent variations, contributing to underfitting.
\ARR{It's likely that the estimators learned smoothed trends associated with available environmental inputs. The addition of lagged weather inputs led to better estimation of day/night peak/trough features,} reflecting the time required for the fiber's ambient conditions to respond to the environmental changes. {This may be the result of} additional material that may be on the fiber that needs to first equilibrate with the environmental conditions. \sout{(e.g., dirt, sheath, fiber huts, sleet).}

\sout{Additionally it is also possible that the model is using the history $\mathcal{F}_X$ to filter out the effect of sharp sub-peaks that vary on a much shorter timescale than $X$. Without this inclusion, the model was likely overfitting to the noise, leading to poorer test RMSLE.
Our choice of estimator parameters for our random forest estimator, see Section \ref{sec: ML_methods}, effectively introduced a strong averaging effect over the noise distribution, and possibly over the deterministic signal. We note that as a result of the low complexity of decision trees in the forest, there was a strong possibility of underfitting contributing to increased bias in the random forest estimator. Increasing the complexity of the estimator would require increasing the training dataset size to prevent overfitting to noise.
When analyzing the log residuals of the estimator, we saw neighborhoods of positive residuals and neighborhoods of negative residuals, but with the former of comparatively weaker magnitude than the latter (see Fig. 11 in Appendix A). This is sensible as our estimator consistently overestimated the nightly quiet activity troughs to a larger proportional degree as compared to how it underestimated the daytime peaks. This also resulted in inflated RMSLE scores due to worse relative nighttime estimation.}

\sout{There are also avenues we could take to further reduce errors without utilizing more complex estimation methods.
Addition of weather variables such as precipitation, solar intensity, and wind gusts could improve estimator performance, and could correct for the ambiguity at 100\% $H$. We also note that the geographic region of the fiber network is large enough that local weather differences over two-minute intervals could be nontrivial. Our infrastructure is within the effective radius of three weather stations and utilizing aggregate weather metrics over the region could further reduce RMSLE errors. Furthermore, we have not taken into account climate patterns (e.g the El Nino Southern Oscillation and polar vortex) which can result in polarization drift events that cannot be explained by just our environmental feature set. Additional data and inclusion of inputs that quantify these patterns, specific to geographical regions, could improve estimation performance. Fundamentally, if we had measurements on the environment conditions at different segments in the fiber rather than at weather stations, we could develop an estimator built around the theoretical model detailed in Section \ref{sec: phys_model} and Appendix B.}

\ARR{It's strongly likely that a significant amount of variability in the target dataset is uncaptured by using only weather inputs. Unmonitored environmental factors and spatial variations along the fiber may account for the remaining uncaptured variability.} Addition of weather variables such as precipitation, solar intensity, and wind gusts could improve estimator performance, and measurements from multiple weathering monitoring sites could provide a better characterization of ambient conditions across the entire fiber connection. \ARR{Heavy vehicular traffic, machinery, and indoor activities were not monitored and may also contributed to unexplained variability. Lastly, local environmental measurements along the fiber connection would enable us to develop an estimator built around a rigorous multi-segment fiber model (see Supplemental 1).}



\sout{With regard to interpreting feature contribution, the presence of strong multicollinearity in the input feature set naturally lead to the regressors arbitrarily assigning regression weights among
the correlated feature inputs, making interpretation of the regression weights highly unreliable. We note that the correlation scores (Pearson’s and Spearman’s) used in this work do not estimate
the proportional strength of the dependencies among a set of features. In the case of Pearson’s correlation, we only measure how linear the trend is. In the case of Spearman’s correlation, we measure how monotonic the trend is. We saw that the temperature and relative humidity had similar Spearman’s correlation with the spectral area, see Fig. 7, while exhibiting strong Pearson’s correlation with each other on daily timescales. However, this does not imply that changes in one of the inputs leads to the same change in the spectral area. We know that the thermo-optic and thermal expansion coefficients are much larger than the equivalents due to humidity, but the ML model cannot infer this. Calculating the feature contribution with strongly correlated features therefore is very challenging as highly correlated features can serve as proxies of each other in regression models and it is challenging to select the regression model that best reflects reality without knowledge of the physical mechanisms} 

With regard to interpreting feature contribution, the presence of strong multicollinearity in the input features limited interpretation of feature contributions in the estimators. We found limited success in extracting reliable results using xAI methods, including SHAP (Shapley explanative scores) and ALE (accumulated local effects)~\cite{krishna2024disagreement}. 
Lab-controlled measurements would help isolate these effects and assess physical contributions.

\subsection{Ongoing work}
In the future, we will be extending the analysis beyond the $Y_{\text{Hz}}$ components to $Y_\text{kHz}$ and $Y_\text{MHz}$, where we 
expect different environmental dependencies, particularly wind gusts\sout{(for the higher frequency drifts [38])}~\cite{LIU201928}. Our lead objective is to obtain a complete characterization of polarization drift, using weather, in the frequency domain.
We also plan to evaluate the possibility of short-time forecasting using lagged weather and polarization drift histories\sout{, and to possibly improve network stability and synchronization}. 
Additional data collection is ongoing and will allow us to further test the utility of ML estimators for polarization drift characterization. \sout{In the future, }We also seek to develop a more formal analysis of the estimator performance using statistical learning theory~\cite{Vapnik95}, and evaluate other ML estimator methods for generalization performance. 

\subsection{Applications for quantum networking}
The results are relevant to quantum/optical networking systems that distribute polarization-entangled photons over fiber, where polarization drift degrades entanglement throughput and fidelity. \sout{Since classical and quantum electromagnetic fields interact similarly with fiber optics, there is direct applicability towards quantum network testbeds leveraging polarization encoded photons.}
In addition, our framework for estimating the features $\mathcal{Y}$ of the relative polarization power enables weather-aware prediction of SOP-drift features from \sout{meteorological and other} environmental telemetry. 
The estimated low frequency components can be useful in approximating a mean quantum link-state signal that can be mapped to polarization-control parameters (e.g bandwidth, tuning parameters) to stabilize throughput throughout the year. 
Consider the case of a PID control loop for polarization stabilization with a Stokes measurement as the error signal~\cite{Chapman:24}. We could determine a mapping between the $\mathcal{Y}$ features over different frequency ranges to the PID parameters, enabling continuous optimized tuning and nonlinear corrections. 

Additionally, link-state estimates {could possibly} be converted into link-quality metrics (e.g. entanglement throughput and fidelity~\cite{rao_2025_ethcap, JC23_qpt}) and used by the control plane for proactive scheduling and routing\sout{: for example, rescheduling entanglement generation}.
Our estimators could possibly support link-health classification, allowing network operators to distinguish environmental effects from equipment \sout{damage or instrument drift}issues~\cite{RODE2025104047, Charlton:17}.\sout{In practice, we would still require a polarization compensation system to carry out robust entanglement distribution on the network.}
While\sout{ results in the literature} previous research has shown partially robust entanglement distribution on \sout{specific} aerial and buried fiber networks~\cite{sena2025highfidelityquantumentanglementdistribution}, the extension of state-of-art quantum networking solutions to fiber infrastructure located in more challenging environments\sout{ requiring high control bandwidth} is still under study. 

Finally, we note that although this work considers characterization of sub-Hz frequency components of polarization drift, the framework developed here is readily extendable for higher frequency components. {The results carry} insights on developing stable quantum links, link state estimators, and polarization compensation systems on deployed fiber.

\begin{backmatter}
\bmsection{Acknowledgment}
This research is performed at Oak Ridge National Laboratory, managed by UT-Battelle, LLC, for the U.S. Department of Energy under contract no. DE-AC05-00OR22725.
U.S. Department of Energy, Office of Science, Advanced Scientific Computing Research, under the  PiQSci: Performance Integrated Quantum Scalable Internet (Field Work Proposal ERKJ432).

\medskip

\noindent The authors declare no conflicts of interest

\bmsection{Data Availability Statement}
Data for the numerical results, ML training data, and model presented in this paper are not publicly available at this time but may be obtained from the authors upon reasonable request.

\bmsection{Supplemental document}
See Supplemental 1 for supporting content.
\end{backmatter}

\appendix

\bibliography{bibgr.bib}

@article{UlrichPM,
    author = {Ulrich, R.},
    title = {Polarization stabilization on single‐mode fiber},
    journal = {Applied Physics Letters},
    volume = {35},
    number = {11},
    pages = {840-842},
    year = {1979},
    month = {12},
    issn = {0003-6951},
    doi = {10.1063/1.90999},
    url = {https://doi.org/10.1063/1.90999},
    eprint = {https://pubs.aip.org/aip/apl/article-pdf/35/11/840/18440236/840_1_online.pdf},
}

@article{LOPEZ2026344838,
title = {The importance of choosing a proper validation strategy in predictive models. Part 2: Recipes for (avoiding) overfitting-A tutorial},
journal = {Analytica Chimica Acta},
volume = {1384},
pages = {344838},
year = {2026},
issn = {0003-2670},
doi = {https://doi.org/10.1016/j.aca.2025.344838},
url = {https://www.sciencedirect.com/science/article/pii/S0003267025012322},
author = {Eneko Lopez and Giulia Gorla and Jaione Etxebarria-Elezgarai and Jose Manuel Amigo and Andreas Seifert}
}

@book{hastie09elements,
  title = {The Elements of Statistical Learning: Data Mining, Inference, and Prediction},
  author = {Trevor Hastie and Robert Tibshirani and Jerome Friedman},
  edition = {2nd},
  year = {2009},
  publisher = {Springer-Verlag},
  address = {},
  isbn = {978-0-387-84857-0},
  url = {https://link.springer.com/book/10.1007/978-0-387-84858-7}
}

@article{Martinelli:06,
author = {Mario Martinelli and Paolo Martelli and Silvia Maria Pietralunga},
journal = {J. Lightwave Technol.},
number = {11},
pages = {4172--4183},
publisher = {Optica Publishing Group},
title = {Polarization Stabilization in Optical Communications Systems},
volume = {24},
month = {Nov},
year = {2006},
url = {https://opg.optica.org/jlt/abstract.cfm?URI=jlt-24-11-4172},
}

@article{czegledi2016polarization,
  title={Polarization drift channel model for coherent fibre-optic systems},
  author={Czegledi, Cristian B and Karlsson, Magnus and Agrell, Erik and Johannisson, Pontus},
  journal={Scientific reports},
  volume={6},
  number={1},
  pages={21217},
  year={2016},
  publisher={Nature Publishing Group UK London}
}

@article{Wai:94bireffiber,
author = {P. K. A. Wai and C. R. Menyuk},
journal = {Opt. Lett.},
number = {19},
pages = {1517--1519},
publisher = {Optica Publishing Group},
title = {Polarization decorrelation in optical fibers with randomly varying birefringence},
volume = {19},
month = {Oct},
year = {1994},
url = {https://opg.optica.org/ol/abstract.cfm?URI=ol-19-19-1517},
doi = {10.1364/OL.19.001517},
}

@article{Chapman:24,
author = {Joseph C. Chapman and Muneer Alshowkan and Kazi Reaz and Tian Li and Mariam Kiran},
journal = {Opt. Express},
number = {26},
pages = {47589--47619},
publisher = {Optica Publishing Group},
title = {Continuous automatic polarization channel stabilization from heterodyne detection of coexisting dim reference signals},
volume = {32},
month = {Dec},
year = {2024},
url = {https://opg.optica.org/oe/abstract.cfm?URI=oe-32-26-47589},
doi = {10.1364/OE.543704},
}

@article{Craddock24Qu,
  title = {Automated Distribution of Polarization-Entangled Photons Using Deployed New York City Fibers},
  author = {Craddock, Alexander N. and Lazenby, Anne and Portmann, Gabriel Bello and Sekelsky, Rourke and Flament, Mael and Namazi, Mehdi},
  journal = {PRX Quantum},
  volume = {5},
  issue = {3},
  pages = {030330},
  numpages = {7},
  year = {2024},
  month = {Aug},
  publisher = {American Physical Society},
  doi = {10.1103/PRXQuantum.5.030330},
  url = {https://link.aps.org/doi/10.1103/PRXQuantum.5.030330}
}

@article{KarlssonPMD,
author = {Karlsson, O. and Brentel, Jonas and Andrekson, Peter},
year = {2000},
month = {08},
pages = {941-951},
title = {Long-Term Measurement of PMD and Polarization Drift in Installed Fibers},
volume = {18},
journal = {Lightwave Technology, Journal of},
doi = {10.1109/50.850739}
}

@article{sena2025highfidelityquantumentanglementdistribution,
author = {Matheus Sena and Mael Flament and Shane Andrewski and Ioannis Caltzidis and Niccol\`{o} Bigagli and Thomas Rieser and Gabriel Bello Portmann and Rourke Sekelsky and Ralf-Peter Braun and Alexander N. Craddock and Maximilian Schulz and Klaus D. J\"{o}ns and Michaela Ritter and Marc Geitz and Oliver Holschke and Mehdi Namazi},
journal = {J. Opt. Commun. Netw.},
number = {12},
pages = {1072--1081},
publisher = {Optica Publishing Group},
title = {High-fidelity quantum entanglement distribution in metropolitan fiber networks with co-propagating classical traffic},
volume = {17},
month = {Dec},
year = {2025},
url = {https://opg.optica.org/jocn/abstract.cfm?URI=jocn-17-12-1072},
doi = {10.1364/JOCN.575396},
}

@inproceedings{Eastman:25,
  author    = {Eastman, Ely and Liu, Hyouin and Ramesh, Anirudh and Chung, Joaquin and Chitambar, Eric and Kettimuthu, Rajkumar and Kumar, Prem},
  title     = {Direct Tracking of State-of-Polarization Fluctuations in Fiber due to Polarization Mode Dispersion Using Machine Learning},
  booktitle = {Frontiers in Optics + Laser Science 2025 (FiO, LS)},
  year      = {2025},
  paper     = {JW5A.9},
  doi       = {10.1364/fio.2025.jw5a.9},
  publisher = {Optica Publishing Group}
}

@article{LIU201928,
title = {Analysis of polarization fluctuation in long-distance aerial fiber for QKD system design},
journal = {Optical Fiber Technology},
volume = {48},
pages = {28-33},
year = {2019},
issn = {1068-5200},
doi = {https://doi.org/10.1016/j.yofte.2018.12.012},
url = {https://www.sciencedirect.com/science/article/pii/S1068520018303183},
author = {Rende Liu and Hao Yu and Jiye Zan and Song Gao and Liwei Wang and Mulan Xu and Jun Tao and Jianhong Liu and Qing Chen and Yong Zhao}
}

@article{Poole:91,
author = {C. D. Poole and J. H. Winters and J. A. Nagel},
journal = {Opt. Lett.},
number = {6},
pages = {372--374},
publisher = {Optica Publishing Group},
title = {Dynamical equation for polarization dispersion},
volume = {16},
month = {Mar},
year = {1991},
url = {https://opg.optica.org/ol/abstract.cfm?URI=ol-16-6-372},
doi = {10.1364/OL.16.000372},
}

@article{KhouloudNokiaANN,
author = {Abdelli, Khouloud and Lonardi, Matteo and Gripp, Jurgen and Olsson, Samuel and Boitier, Fabien and Layec, Patricia},
year = {2025},
month = {07},
pages = {1-12},
title = {Forecasting of Weather-Induced State of Polarization Changes in Aerial Fibers},
volume = {PP},
journal = {Journal of Lightwave Technology},
doi = {10.1109/JLT.2025.3549593}
}

@article{alshowkan21testbed,
  title = {Reconfigurable Quantum Local Area Network Over Deployed Fiber},
  author = {Alshowkan, Muneer and Williams, Brian P. and Evans, Philip G. and Rao, Nageswara S.V. and Simmerman, Emma M. and Lu, Hsuan-Hao and Lingaraju, Navin B. and Weiner, Andrew M. and Marvinney, Claire E. and Pai, Yun-Yi and Lawrie, Benjamin J. and Peters, Nicholas A. and Lukens, Joseph M.},
  journal = {PRX Quantum},
  volume = {2},
  issue = {4},
  pages = {040304},
  numpages = {13},
  year = {2021},
  month = {Oct},
  publisher = {American Physical Society},
  doi = {10.1103/PRXQuantum.2.040304},
  url = {https://link.aps.org/doi/10.1103/PRXQuantum.2.040304}
}

@article{
rao_2025_ethcap,
author = {Nageswara Rao  and Muneer Alshowkan  and Aneesh Ramaswamy  and Joseph C. Chapman  and Nicholas A. Peters  and Hsuan-Hao Lu  and Joseph M. Lukens  and Saikat Guha },
title = {Entanglement Throughput Over Fiber Connections: Measurements and Capacity Estimates},
journal = {TechRxiv},
volume = {2025},
number = {0909},
pages = {},
year = {2025},
doi = {10.36227/techrxiv.175745398.85881596/v1},
URL = {https://www.techrxiv.org/doi/abs/10.36227/techrxiv.175745398.85881596/v1},
eprint = {https://www.techrxiv.org/doi/pdf/10.36227/techrxiv.175745398.85881596/v1}
}

@article{RODE2025104047,
title = {Machine learning opportunities for integrated polarization sensing and communication in optical fibers},
journal = {Optical Fiber Technology},
volume = {90},
pages = {104047},
year = {2025},
issn = {1068-5200},
doi = {https://doi.org/10.1016/j.yofte.2024.104047},
url = {https://www.sciencedirect.com/science/article/pii/S1068520024003924},
author = {Andrej Rode and Mohammad Farsi and Vincent Lauinger and Magnus Karlsson and Erik Agrell and Laurent Schmalen and Christian Häger}
}

@article{Charlton:17,
author = {Douglas Charlton and Steven Clarke and David Doucet and Maurice O'Sullivan and Daniel L Peterson and Darryl Wilson and Glenn Wellbrock and Michel B\'{e}langer},
journal = {Opt. Express},
number = {9},
pages = {9689--9696},
publisher = {Optica Publishing Group},
title = {Field measurements of SOP transients in OPGW, with time and location correlation to lightning strikes},
volume = {25},
month = {May},
year = {2017},
url = {https://opg.optica.org/oe/abstract.cfm?URI=oe-25-9-9689},
doi = {10.1364/OE.25.009689},
}

@Article{app9061178,
AUTHOR = {Weng, Yi and Wang, Junyi and Pan, Zhongqi},
TITLE = {Recent Advances in DSP Techniques for Mode Division Multiplexing Optical Networks with MIMO Equalization: A Review},
JOURNAL = {Applied Sciences},
VOLUME = {9},
YEAR = {2019},
NUMBER = {6},
pages = {1178},
URL = {https://www.mdpi.com/2076-3417/9/6/1178},
ISSN = {2076-3417},
DOI = {10.3390/app9061178}
}

@article{JC23_qpt,
  title = {Coexistent Quantum Channel Characterization Using Spectrally Resolved Bayesian Quantum Process Tomography},
  author = {Chapman, Joseph C. and Lukens, Joseph M. and Alshowkan, Muneer and Rao, Nageswara and Kirby, Brian T. and Peters, Nicholas A.},
  journal = {Phys. Rev. Appl.},
  volume = {19},
  issue = {4},
  pages = {044026},
  numpages = {18},
  year = {2023},
  month = {Apr},
  publisher = {American Physical Society},
  doi = {10.1103/PhysRevApplied.19.044026},
  url = {https://link.aps.org/doi/10.1103/PhysRevApplied.19.044026}
}

@book{Vapnik95,
  title     = {The Nature of Statistical Learning Theory},
  author    = {Vapnik, Vladimir N.},
  year      = {1999},
  edition   = {2nd},
  publisher = {Springer},
  address   = {},
  series    = {Information Science and Statistics},
  isbn      = {978-0387987804}
}

@article{ProbstRFtuning,
author = {Probst, Philipp and Wright, Marvin N. and Boulesteix, Anne-Laure},
title = {Hyperparameters and tuning strategies for random forest},
journal = {WIREs Data Mining and Knowledge Discovery},
volume = {9},
number = {3},
pages = {1301},
doi = {https://doi.org/10.1002/widm.1301},
url = {https://wires.onlinelibrary.wiley.com/doi/abs/10.1002/widm.1301},
eprint = {https://wires.onlinelibrary.wiley.com/doi/pdf/10.1002/widm.1301},
year = {2019}
}

@article{Biau2015ARF,
  title={A random forest guided tour},
  author={G{\'e}rard Biau and Erwan Scornet},
  journal={TEST},
  year={2015},
  volume={25},
  pages={197 - 227},
  url={https://api.semanticscholar.org/CorpusID:14518730}
}

@article{scikit-learn,
  title={Scikit-learn: Machine Learning in {P}ython},
  author={Pedregosa, F. and Varoquaux, G. and Gramfort, A. and Michel, V.
          and Thirion, B. and Grisel, O. and Blondel, M. and Prettenhofer, P.
          and Weiss, R. and Dubourg, V. and Vanderplas, J. and Passos, A. and
          Cournapeau, D. and Brucher, M. and Perrot, M. and Duchesnay, E.},
  journal={Journal of Machine Learning Research},
  volume={12},
  pages={2825--2830},
  year={2011}
}

@article{krishna2024disagreement,
    title={The Disagreement Problem in Explainable Machine Learning: A Practitioner{\textquoteright}s Perspective},
    author={Satyapriya Krishna and Tessa Han and Alex Gu and Steven Wu and Shahin Jabbari and Himabindu Lakkaraju},
    journal={Transactions on Machine Learning Research},
    issn={2835-8856},
    year={2024},
    url={https://openreview.net/forum?id=jESY2WTZCe},
    note={}
}

@article{Barcik:20,
author = {Peter Barcik and Petr Munster},
journal = {Opt. Express},
number = {10},
pages = {15250--15257},
publisher = {Optica Publishing Group},
title = {Measurement of slow and fast polarization transients on a fiber-optic testbed},
volume = {28},
month = {May},
year = {2020},
url = {https://opg.optica.org/oe/abstract.cfm?URI=oe-28-10-15250},
doi = {10.1364/OE.390649},
}

@article{Woodwardfiber,
author = {Woodward, S.L. and Nelson, Lynn and Schneider, Charles and Knox, Laurie and O'Sullivan, Maurice and Laperle, Charles and Moyer, Michael and Foo, Sik},
year = {2014},
month = {02},
pages = {213-216},
title = {Long-term observation of PMD and SOP on installed fiber routes},
volume = {26},
journal = {Photonics Technology Letters, IEEE},
doi = {10.1109/LPT.2013.2290473}
}

@inproceedings{Bohata,
  author    = {Bohata, J. and Zvanovec, S. and Pisarik, M.},
  title     = {Outdoor atmospheric influence on polarization mode dispersion in optical cables},
  booktitle = {General Assembly and Scientific Symposium (URSI GASS)},
  year      = {2014},
  pages     = {1--4},
  doi       = {10.1109/URSIGASS.2014.6929421}
}

@INPROCEEDINGS{Tremblay,
  author={Tremblay, Christine and Michel, Annie and Tanoh, Marie Janvier and Bélanger, Michel P. and Clarke, Steven and Charlton, Douglas W. and Peterson, Daniel L. and Wellbrock, Glenn A.},
  booktitle={2017 19th International Conference on Transparent Optical Networks (ICTON)}, 
  title={Dynamics of polarization fluctuations in aerial and buried links}, 
  year={2017},
  volume={},
  number={},
  pages={1-1},
  doi={10.1109/ICTON.2017.8024999}}

@techreport{EXFO,
  author      = {EXFO},
  title       = {Case study: {SOP} and {PMD} Measurements on Aerial Fiber Under Wind-Induced Oscillations and Vibrations},
  institution = {EXFO},
  type        = {Technical Note},
  number      = {TN039},
  url         = {https://www.exfo.com/contentassets/1f4366a69b53491ba6aab63019efe455/exfo_tnote040_sop-pmd-measurement-case-study_en.pdf},
  note        = {}
}

@article{bifrost,
  title = {bifrost: A first-principles model of polarization mode dispersion in optical fiber},
  author = {Banner, Patrick R. and Rolston, S. L. and Britton, Joseph W.},
  journal = {Phys. Rev. Appl.},
  volume = {25},
  issue = {3},
  pages = {034054},
  numpages = {20},
  year = {2026},
  month = {Mar},
  publisher = {American Physical Society},
  doi = {10.1103/xgqr-rlmf},
  url = {https://link.aps.org/doi/10.1103/xgqr-rlmf}
}

@Article{wangphotonics10020103,
AUTHOR = {Wang, Lidong and Liao, Meisong and Yu, Fei and Li, Weichang and Xu, Jiacheng and Hu, Lili and Gao, Weiqing},
TITLE = {Thermal Sensitivity of Birefringence in Polarization-Maintaining Hollow-Core Photonic Bandgap Fibers},
JOURNAL = {Photonics},
VOLUME = {10},
YEAR = {2023},
NUMBER = {2},
pages = {103},
URL = {https://www.mdpi.com/2304-6732/10/2/103},
ISSN = {2304-6732},
DOI = {10.3390/photonics10020103}
}

@article{rodgers1988thirteen,
  title={Thirteen ways to look at the correlation coefficient},
  author={Rodgers, Joseph Lee and Nicewander, W Alan},
  journal={The American Statistician},
  volume={42},
  number={1},
  pages={59--66},
  year={1988},
  publisher={Taylor \& Francis}
}

@article{PCAreview,
    author = {Jolliffe, Ian T. and Cadima, Jorge},
    title = {Principal component analysis: a review and recent developments},
    journal = {Philosophical Transactions of the Royal Society A: Mathematical, Physical and Engineering Sciences},
    volume = {374},
    number = {2065},
    pages = {20150202},
    year = {2016},
    month = {04},
    issn = {1364-503X},
    doi = {10.1098/rsta.2015.0202},
    url = {https://doi.org/10.1098/rsta.2015.0202},
    eprint = {https://royalsocietypublishing.org/rsta/article-pdf/doi/10.1098/rsta.2015.0202/1381479/rsta.2015.0202.pdf},
}

@article{Brodsky_2006_PMD,
  title = {Polarization Mode Dispersion of Installed Fibers},
  author = {Brodsky, Misha and Frigo, Nicholas J. and Boroditsky, Misha and Tur, Moshe},
  journal = {Journal of Lightwave Technology},
  volume = {24},
  number = {12},
  pages = {4584--4599},
  year = {2006},
  publisher = {}
}






\end{document}